\documentclass[longauth]{aa} 

\usepackage{graphicx}

\usepackage{txfonts}
\usepackage{longtable}
\usepackage{ccaption}
\captionstyle{\raggedright}
\usepackage{soul}

\usepackage{pdflscape}

\usepackage[dvipsnames]{xcolor}
\newcommand{\kms}{km\,s$^{-1}$\xspace}

\newcommand{\pvb}{SN~2020pvb\xspace}

\graphicspath{ {./figures/} }

\begin{document}

   \title{\pvb: a Type IIn-P supernova with a precursor outburst}

   \author{N.~Elias-Rosa\inst{1,2}
          \and S.~J.~Brennan\inst{3,4}
          \and S.~Benetti\inst{1}
          \and E.~Cappellaro\inst{1}
          \and A.~Pastorello\inst{1}
          \and A.~Kozyreva\inst{5}
          \and P.~Lundqvist\inst{3}
          \and M.~Fraser\inst{4}
          \and J.~P. Anderson\inst{6,7}  
          \and Y.-Z.~Cai\inst{8,9,10}
          \and T.-W.~Chen\inst{11}          
          \and M.~Dennefeld \inst{12}
          \and M.~Gromadzki\inst{13} 
          \and C.~P.~Guti\'errez\inst{2,14}    
          \and N.~Ihanec\inst{13}
          \and C.~Inserra\inst{15} 
          \and E.~Kankare\inst{16,17}
          \and R.~Kotak\inst{16}
          \and S.~Mattila\inst{16,18}
          \and S.~Moran\inst{16}
          \and T.~E.~M\"uller-Bravo\inst{2,14}
          \and P.~J.~Pessi\inst{3}. 
          \and G.~Pignata\inst{19} 
          \and A.~Reguitti\inst{20,1}
          \and T.~M.~Reynolds\inst{21,16}
          \and S.~J.~Smartt\inst{22,23}
          \and K.~Smith\inst{23}
          \and L.~Tartaglia\inst{24}
          \and G.~Valerin\inst{1}
          \and T.~de Boer\inst{25}
          \and K.~Chambers\inst{25}
          \and A.~Gal-Yam\inst{26}
          \and H.~Gao\inst{25}          
          \and S.~Geier\inst{27,28}
          \and P.~A.~Mazzali\inst{29,30}
          \and M.~Nicholl\inst{23}
          \and F.~Ragosta\inst{31,32}
          \and A.~Rest\inst{33,34}
          \and O.~Yaron\inst{26}
          \and D.~R.~Young\inst{23} 
          }

   \institute{INAF -- Osservatorio Astronomico di Padova, vicolo dell'Osservatorio 5, Padova I-35122, Italy\\ 
              \email{nancy.elias@inaf.it}
    \and 
    Institute of Space Sciences (ICE, CSIC), Campus UAB, Carrer de Can Magrans s/n, E-08193 Barcelona, Spain
    \and 
    The Oskar Klein Centre, Department of Astronomy, Stockholm University, AlbaNova, SE-10691 Stockholm, Sweden
    \and 
    School of Physics, O’Brien Centre for Science North, University College Dublin, Belfield, Dublin 4, Ireland
    \and 
    Heidelberger Institut f\"ur Theoretische Studien, Schloss-Wolfsbrunnenweg 35, 69118 Heidelberg, Germany
    \and 
    European Southern Observatory, Alonso de C\'ordova 3107, Casilla 19, Santiago, Chile
    \and 
    Millennium Institute of Astrophysics (MAS), Nuncio Monse\~{n}or S\'otero Sanz 100, Off. 104, Providencia, Santiago, Chile
    \and 
    Yunnan Observatories, Chinese Academy of Sciences, Kunming 650216, PR China
    \and 
    Key Laboratory for the Structure and Evolution of Celestial Objects, Chinese Academy of Sciences, Kunming 650216, PR China
    \and 
    International Centre of Supernovae, Yunnan Key Laboratory, Kunming 650216, P.R. China
    \and 
    Graduate Institute of Astronomy, National Central University, 300 Jhongda Road, 32001 Jhongli, Taiwan
    \and 
    CNRS, Institut d'Astrophysique de Paris (IAP) and Sorbonne Universit\'e (Paris 6), 98bis, Boulevard Arago F-75014, Paris
    \and 
    Astronomical Observatory, University of Warsaw, Al. Ujazdowskie 4, 00-478 Warszawa, Poland
    \and 
    Institut d’Estudis Espacials de Catalunya (IEEC), E-08034 Barcelona, Spain.
    \and 
    Cardiff Hub for Astrophysics Research and Technology, School of Physics \& Astronomy, Cardiff University, Queens Buildings, The Parade, Cardiff, CF24 3AA, UK
    \and 
    Tuorla Observatory, Department of Physics and Astronomy, FI-20014 University of Turku, Finland
    \and 
    Turku Collegium for Science, Medicine and Technology, University of Turku, FI-20014 Turku, Finland
    \and 
    School of Sciences, European University Cyprus, Diogenes street, Engomi, 1516 Nicosia, Cyprus.
    \and 
    Instituto de Alta Investigaci\'on, Universidad de Tarapac\'a, Casilla 7D, Arica, Chile 
    \and 
    INAF, Osservatorio Astronomico di Brera, Via E. Bianchi 46, I-23807, Merate (LC), Italy
    \and 
    Cosmic DAWN centre, Niels Bohr Institute, University of Copenhagen, Jagtvej 128, 2200 K{\o}benhavn N, Denmark
    \and 
    Department of Physics, University of Oxford, Denys Wilkinson Building, Keble Road, Oxford, OX1 3RH, UK
    \and 
    Astrophysics Research Centre, School of Mathematics and Physics, Queen’s University Belfast, Belfast BT7 1NN, UK
    \and 
    INAF -- Osservatorio Astronomico d'Abruzzo, via M. Maggini snc, Teramo, I-64100, Italy
    \and 
    Institute for Astronomy, University of Hawaii, 2680 Woodlawn Drive, Honolulu HI 96822, USA
    \and 
    Department of particle physics and astrophysics, Weizmann Institute of Science, 76100 Rehovot, Israel
    \and 
    Gran Telescopio Canarias (GRANTECAN), Cuesta de San Jos\'e s/n, 38712 Bre\~na Baja, La Palma, Spain
    \and 
    Instituto de Astrofísica de Canarias, V\'ia L\'actea s/n, 38200 La Laguna, Tenerife, Spain
    \and 
    Astrophysics Research Institute, Liverpool John Moores University, ic2, 146 Brownlow Hill, Liverpool L3 5RF, UK
    \and 
    Max-Planck Institut f\"ur Astrophysik, Karl-Schwarzschild-Str. 1, D-85741 Garching bei M\"{u}nchen, Germany
    \and 
    INAF, Osservatorio Astronomico di Roma, via Frascati 33, I-00078 Monte Porzio Catone (RM), Italy
    \and 
    Space Science Data Center – ASI, Via del Politecnico SNC, 00133 Roma, Italy
    \and 
    Department of Physics and Astronomy, Johns Hopkins University, 3400 North Charles Street, Baltimore, MD 21218, USA
    \and 
    Space Telescope Science Institute, 3700 San Martin Drive, Baltimore, MD 21218, USA
}

   \date{Received ..., 2023; accepted ..., 2024}

 
  \abstract
   {We present photometric and spectroscopic data sets for \pvb, a Type IIn-P supernova (SN) similar to SNe~1994W, 2005cl, 2009kn and 2011ht, with a precursor outburst detected (PS1 $w$-band $\sim$ -13.8 mag) around four months before the $B$-band maximum light. \pvb presents a relatively bright light curve peaking at $M_B$ = -17.95 $\pm$ 0.30 mag and a plateau lasting at least 40 days before it went in solar conjunction. After this, the object is no longer visible at phases $>$ 150 days above -12.5 mag in the $B$-band, suggesting that the \pvb ejecta interacts with a dense spatially confined circumstellar envelope. \pvb shows in its spectra strong Balmer lines and a forest of \ion{Fe}{ii} lines with narrow P Cygni profiles. Using archival images from the Hubble Space Telescope, we constrain the progenitor of \pvb to have a luminosity of $\log(L/L_{\sun}) \lesssim$ 5.4, ruling out any single star progenitor over 50 $M_\sun$. All in all, \pvb is a Type IIn-P whose progenitor star had an outburst $\sim$ 0.5 yr before the final explosion, the material lost during this outburst is probably playing a role in shaping the physical properties of the supernova.
   }


   \maketitle
%
\section{Introduction}\label{SNintro}

Massive stars ($\geq$ 8 $M_{\sun}$) can lose mass via steady winds, binary interaction, or (more rarely) as a consequence of dramatic eruptions, generating a dense and often structured circumstellar medium \citep[CSM;][]{Weis2001,Vink2008,Smith2011b,Smith2017}. In the case of single massive stars, non-terminal outbursts that can produce this CSM are instabilities occurring when the star approaches the end of its evolution. When such a massive star explodes as a supernova (SN), the ejected material interacts with the H-rich CSM. As a consequence, the resulting transient spectra present a blue continuum with superposed narrow emission lines (with inferred full-width-at-half-maximum (FWHM) velocities from a few tens to a few hundred \kms), arising from the CSM excited by the shock interaction emission. The interaction can mask the innermost ejecta (as well as the explosion mechanism; \citealt{chevalier94}). However, if the CSM is optically thin or with a particular geometric configuration, it is also possible to detect high-velocity components (a few 10$^{3}$ \kms) arising from the SN ejecta. These SNe are known as Type IIn SNe \citep[named after the detection of narrow emission lines in their spectra; e.g.][]{Schlegel1990}. Despite the similarity of their spectroscopic properties, SNe IIn light curves are quite heterogeneous, showing both slow and fast declining SNe, as well as faint ($M_R \lesssim$ -17 mag) and very bright ($M_R \gtrsim$ -19 mag) objects \citep[e.g.][]{kiewe12,taddia13,nyholm20}. 

Type IIn-P SNe (\citealt{mauerhan13b}; see also \citealp{smith17,fraser20}) are a subclass of SNe IIn that exhibit spectra with narrow emission lines throughout their evolution but have light curves with a well-defined plateau mainly in the optical-red and near-infrared (NIR) bands  
(as in Type IIP SNe), followed by an abrupt drop (several magnitudes) onto a radioactive decay tail. Their post-plateau decay also suggests low ejected $^{56}$Ni masses and a low-energy explosion. 
The spectra of Type IIn-P SNe are characterized by a blue continuum and narrow Balmer emission lines together with a forest of narrow P Cygni \ion{Fe}{ii} lines, with a lack of forbidden and high-ionization lines. Specifically, narrow Balmer lines are visible during the earlier phase and are persistent for weeks or months, unlike those observed in very early Type II SNe \citep[flash-ionization features; e.g.][]{khazov16,bruch21,tartaglia21}. SN~1994W \citep{tsvetkov95,cumming97,sollerman98} is the prototype of the Type IIn-P SNe, while other well studied cases are SNe 2009kn \citep{kankare12}, 2005cl \citep{kiewe12} and 2011ht \citep{roming12,mauerhan13b}. Note that \citet{smith13} proposed that the Crab Nebula may be the remnant of one event of this type, albeit with some uncertainties. 

SNe IIn are suggested to originate from high luminosity, massive progenitors \citep[e.g.][]{galyam07,eliasrosa18}, with cases of pre-SN precursor stars variability ($\sim$ 25$\%$ of the SNe IIn cases according to \citealt{strotjohann21}; e.g. \citealt{ofek13b,mauerhan13a,fraser13,ofek14,margutti14,tartaglia16,eliasrosa16,ofek16,thone17,pastorello18,reguitti19}). On the other hand, Type IIn-P have been suggested to originate from the core collapse of intermediate-mass (8-10 $M_{\sun}$) stars \citep[e.g.][]{sollerman98}, where a pre-SN precursor has been observed in only one event, SN 2011ht \citep{fraser13}. Alternative scenarios for SNe~IIn-P include an electron capture SN explosion from a super-AGB (asymptotic giant branch) star \citep{mauerhan13b,smith13}, a non-terminal outburst \citep{dessart09,humphreys12}, a post-merger event \citep{pastorello19}, or even from eruptive outbursts of lower-mass red supergiants \citep[RSGs;][]{chunhui22}. Regardless, this type of event does not seem to be related to very high-mass stars \citep[e.g.][]{smith13,chugai16}. \\

Although ongoing surveys have significantly increased the available sample of SNe, the number of Type IIn-P SNe remains small, limiting our understanding of these objects. Here we present the case of \pvb, a member of this family of SNe, and the second with a precursor outburst detected around four months before the discovery. In the next section (Sect. \ref{SNhostgx}), we provide a summary of the properties of \pvb and its host galaxy. Photometric and spectroscopic data are analysed in Sect. \ref{SNdata}, and the results are shown in Sect. \ref{SNresults}. We discuss the nature of the progenitor star in Sect. \ref{SNprog}, and the nature of the SN in Sect. \ref{SNdiscussion}. 

\section{\pvb: discovery, explosion, distance, and reddening}\label{SNhostgx}

AT~2020pvb ($\alpha$ = 20$^{\rm h}$ 53$^{\rm m}$ 53${\fs}$03, \mbox{$\delta$ = -25$^{\circ}$ 28$\arcmin$ 26${\farcs}$1}; J2000.0)\footnote{\pvb is also known as ATLAS20zmy, PS20gas, and ZTF20acghodf.} was discovered on 2020 July 18.42 UT (MJD = 59048.42) by the Pan-STARRS survey \citep{chambers20_discovery,chambers16,magnier16}, with a PS1 $w$-band magnitude of $w$ = 21.04 $\pm$ 0.18. It was followed by non-detections from different surveys until new and independent discoveries were reported several weeks later by the Asteroid Terrestrial-impact Last Alert System \citep[ATLAS;][]{tonry18,smith20} and the Zwicky Transient Facility \citep[ZTF;][]{masci19}, indicating that the July discovery was rather a pre-SN outburst. The transient was finally classified by ZTF \citep{perley20_class} as a Type IIn SN on 2020 October 15.07 UT.

The first detection of \pvb on the main rise of its supernova-like light curve was obtained on 2020 September 07 (MJD = 59099.37, $o = 19.28 \pm 0.43$ mag) by ATLAS, while the last non-detection was obtained two days before (MJD = 59097.39, $o > 19.8$ mag). We hence estimate the explosion epoch to be on MJD = 59098.38 $\pm$ 1.0 (2020 September 06).

\begin{figure}[!ht]
\centering
\includegraphics[width=0.9\columnwidth]{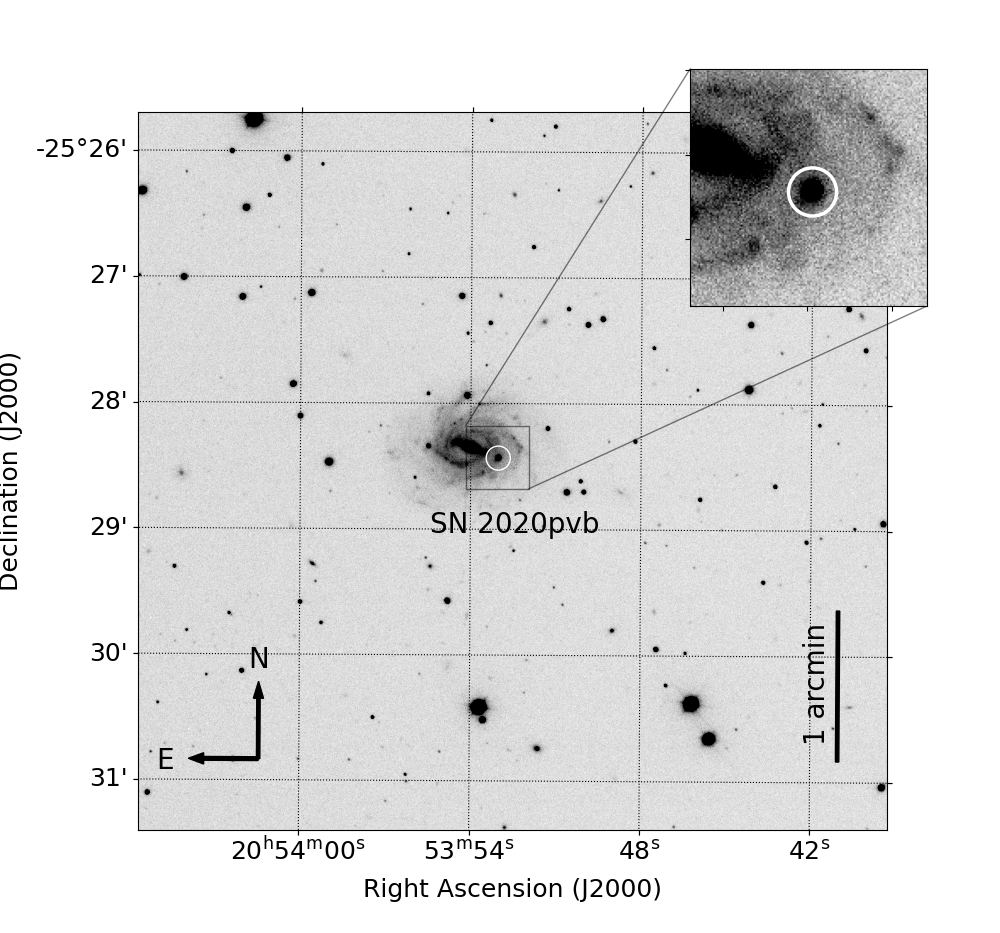}
\caption{Nordic Optical Telescope (NOT) + Alhambra Faint Object Spectrograph and Camera (ALFOSC) image of the \pvb field in the $r$-band obtained on 2020 October 15. The position of the SN is indicated with a circle, and a close-up of the area around the SN is shown in the upper-right inset. }
\label{fig_FC}%
\end{figure}

\pvb is hosted in the barred spiral galaxy NGC~6993 (SB(r)cd\footnote{NED, the NASA/IPAC Extragalactic Database. It is funded by the National Aeronautics and Space Administration and operated by the California Institute of Technology; \url{http://nedwww.ipac.caltech.edu/}.}; \citealt{deVaucouleurs91}; see Fig. \ref{fig_FC}). The recessional velocity of the galaxy, corrected for the Local Group infall into the Virgo cluster \citep{mould00} is v$_{Vir}$ = 6074 $\pm$ 9 \kms (z = 0.02025 $\pm$ 0.00003). Assuming H$_0$ = 73.2 $\pm$ 1.3 km\,s$^{-1}$\,Mpc$^{-1}$\, \citep{riess21}, we derive a distance of 83.0 $\pm$ 1.5 Mpc ($\mu$ = 34.6 $\pm$ 0.1 mag). This distance will be adopted throughout this paper. 

The Milky Way (MW) optical-band extinction in the direction of the transient is $A_{V, \rm MW} = 0.187$ mag (NED; \citealt{schlafly11}). We find no evidence of \ion{Na} {i} D absorption at the redshift of the host galaxy in our highest signal-to-noise (S/N) spectra, which would indicate the absence of gas, and hence likely dust, along the line of sight. Given the large dispersion observed in the equivalent width of the \ion{Na} {i} D absorption lines versus $E(B-V)$ plane \citep[e.g.][]{eliasrosa07,poznanski11,phillips13}, we also estimate the optical extinction toward \pvb, based on comparisons of the object’s spectral energy distribution (SED) with those of other Type~IIn SNe with similar light curves or spectra (see Sect. \ref{SNresults}).
We first match the intrinsic $(B-V)_0$ colour curve of \pvb with that of SNe~IIn-P~1994W \citep{tsvetkov95,sollerman98}, 2009kn \citep{kankare12} and 2011ht \citep{roming12,mauerhan13b}, (see Sect. \ref{SNlc}). We compare the optical SED of \pvb\ at -24.3, 1.7 and 31.9 d from the $B$-band maximum with those of the previous comparison SNe at a similar epoch. The SED of the reference SNe were first corrected for redshift and extinction and then scaled to the distance of \pvb. Assuming $R_{\rm V}$ = 3.1 \citep{cardelli89}, we derive an average $A_{V,host}$ = 0.15 mag and thus, we adopt $A_V$ = 0.34 $\pm$ 0.15\footnote{Root-mean-squared (RMS) uncertainty.} mag, i.e., $E(B-V)$ = 0.11 $\pm$ 0.05 mag, as the total extinction toward \pvb.

\section{Observations}\label{SNdata}

\subsection{Ground-based observations}
Optical $BV$, $ugriz$, and NIR $JHK$ images of \pvb were taken using a large number of observing facilities, listed in Table \ref{table_facilities}. After its discovery, the transient was observed for about two months until it went behind the Sun. The telescopes pointed to the field again around two months later, but the SN was no longer visible with ground-based telescopes.

The photometric observations were reduced in the standard fashion with IRAF and various instrument pipelines. We performed photometry on \pvb and sequence stars using the Automated Photometry Of Transients pipeline \citep[{\sc AutoPhOT};][]{Brennan2022}. Point spread function (PSF) photometry was performed using a PSF model built from bright, isolated sources in the image. The optical magnitudes are calibrated against stars in the vicinity of \pvb with known Vega magnitudes from the American Association of Variable Star Observers (AAVSO) Photometric All-Sky Survey (APASS)\footnote{\url{https://www.aavso.org/apass}} and AB magnitudes from the Sloan Digital Sky Survey (SDSS)\footnote{\url{https://www.sdss.org/}}. We lack any sequence data for the $R$- and $I$-band images, and therefore we approximate the SN magnitudes using the Johnson-to-Sloan band transformation relations from \cite{jester2005} (see Tables \ref{table_ph}, \ref{table_SLph}, and \ref{table_NIRph}).

Late-time (phase $>$ 150 d) photometry of \pvb was computed using the template-subtraction technique to remove the background and hence to measure more accurately the SN magnitudes. The template images were obtained with the New Technology Telescope+EFOSC2\footnote{ESO Faint Object Spectrograph and Camera. \url{https://www.eso.org/sci/facilities/lasilla/instruments/efosc.html}} at La Silla Observatory on 2021 September 15 through the extended Public ESO -European Southern Observatory- Spectroscopic Survey for Transient Objects (ePESSTO+) collaboration \citep{smartt15}. Each SN image was registered geometrically and photometrically with its corresponding template using the {\sc AutoPhOT} pipeline. Using a PSF model, we found the flux upper limits using artificial source injection at the position of \pvb, as described in \cite{Brennan2022}.

For the NIR exposures, we also applied a sky background subtraction using the NOTCam QUICKLOOK v2.5 reduction package\footnote{\url{http://www.not.iac.es/instruments/notcam/guide/observe.html}} for the NOT images and custom IDL routines for the CPAPIR images \citep{artigau04}.\\
 
Our dataset also includes photometry from the Pan-STARRS, ATLAS and ZTF wide-field imaging surveys. We retrieved nine epochs of PS1 $w$-band photometry from the Pan-STARRS survey. The field was observed for five months before the SN was discovered. This photometry was obtained from the flux-weighted mean of each epoch in template-subtracted survey images. Two reference images were used, taken on MJD = 56878.55 (2014 August 9) and MJD = 56886.60 (2014 August 17). The final converted AB magnitudes and upper limits (corresponding to 3 times the standard deviation in the background) are listed in Table \ref{table_PSph}.

ATLAS observed the SN field for five years before its discovery in $cyan$ and $orange$ ($c$ and $o$) filters (broadly similar to $g+r$ and $r+i$, respectively). We performed ATLAS forced photometry \citep{smith20} at the site of \pvb using the host-galaxy template-subtraction technique. The reference images were taken on MJD = 58661 (2019 June 27) in ATLAS $c$-band and MJD = 58708 (2019 August 13) in ATLAS $o$-band. We computed a weighted mean of multiple exposures obtained at each epoch and then converted it to an AB magnitude. We present the final photometry in Table \ref{table_ATLASph}.

ZTF photometry of \pvb was measured on $gr$-band template-subtracted images. The measurements were accessed through the forced photometry \citep{masci19} released from the NASA/IPAC Infrared Science Archive\footnote{\url{https://irsa.ipac.caltech.edu/}}. This photometry is included in Table \ref{table_SLph}.\\

Spectroscopic monitoring of \pvb started on 2020 October 12 and lasted two months. We collected 13 optical spectra\footnote{A NIR spectrum was also obtained with the Especrografo Multiobjeto Infra-Rojo (EMIR) at the Gran Telescopio CANARIAS (GTC). However, this spectrum has a low S/N and lacks any obvious SN flux.}. All the spectra were obtained with the slit aligned along the parallactic angle to minimize differential flux losses caused by atmospheric refraction. The spectroscopic observational log can be found in Table \ref{table_spec}\footnote{The spectra will be made public via the Weizmann Interactive Supernova Data Repository (WISeREP; \citealt{yaron12}).}.

The spectra were reduced following standard procedures with {\sc IRAF} routines via the graphical user interface {\sc FOSCGUI}\footnote{{\sc FOSCGUI} is a graphic user interface aimed at extracting SN spectroscopy and photometry obtained with FOSC-like instruments. It was developed by E. Cappellaro. A package description can be found at \url{http://sngroup.oapd.inaf.it/foscgui.html.}} and the PESSTO pipeline \citep{smartt15}.
The two-dimensional frames were corrected for bias and flat-fielded before the extraction of the one-dimensional spectra. 
The one-dimensional optical spectra were then wavelength calibrated by comparison with arc-lamp exposures obtained the same night, while the flux calibration was done using spectra of standard stars. We also verified the wavelength calibration against the bright night-sky emission lines and attempted to remove the strongest telluric absorption bands present in the spectra (in some cases, residuals are still present after the correction). Finally, the flux calibration of the reduced SN spectra was cross-checked against the broad-band photometry, and flux scaled by a constant value when necessary. 

\subsection{Space telescope observations}
The field of \pvb was also observed with the Neil Gehrels Swift Observatory \citep[SWIFT;][]{gehrels04} at four epochs with ultraviolet (UV) and optical filters. The magnitudes of the SN were measured using aperture photometry with the task {\sc uvotsource} included in the UVOT software package\footnote{https://heasarc.gsfc.nasa.gov/docs/software/heasoft/}. The derived magnitudes are listed in Tables \ref{table_UVph} and \ref{table_ph}. \\

The Hubble Space Telescope ({\sl HST}) also observed the site of \pvb with the Wide Field Channel (WFC, $\sim$ 0$\farcs$05 pixel$^{-1}$) of the Advanced Camera for Surveys (ACS) in F606W ($\sim$ $V$-band) on 2017 July 21 (SNAP-14840; PI A. Bellini), corresponding to 1092.5 days (3 years) before the discovery of \pvb.

The photometry of the stellar-like sources located near the transient position was obtained in the {\sc flc} frames (WFC-calibrated and corrected by charge transfer efficiency images) using {\sc Dolphot} \citep{dolphin00,dolphin16} and choosing the drizzled frame as a reference. We used the recommended {\sc Dolphot} parameters setting for ACS/WFC, considering both aperture sizes RAper=4 and RAper=8 for the photometry. We discuss this further in Sect. \ref{SNprog}. 

The field of \pvb was observed by the Galaxy Evolution Explorer (GALEX; \citealt{martin05}) as part of the All Sky Imaging Survey (ASIS) in both the near-UV (NUV;1800-2800~\AA) and far-UV (FUV; 1300-1800~\AA) for a duration of 168~s on 2006 August 31. No point source is detected coincident with the position of the SN in either band. We estimate a 3 $\sigma$ upper limit of 19.7 and 20.4 mag (AB mag) for NUV and FUV, respectively.
\section{Results}\label{SNresults}

\subsection{Light curves}\label{SNlc}

\begin{figure*}[!ht]
\centering
\includegraphics[width=0.7\textwidth]{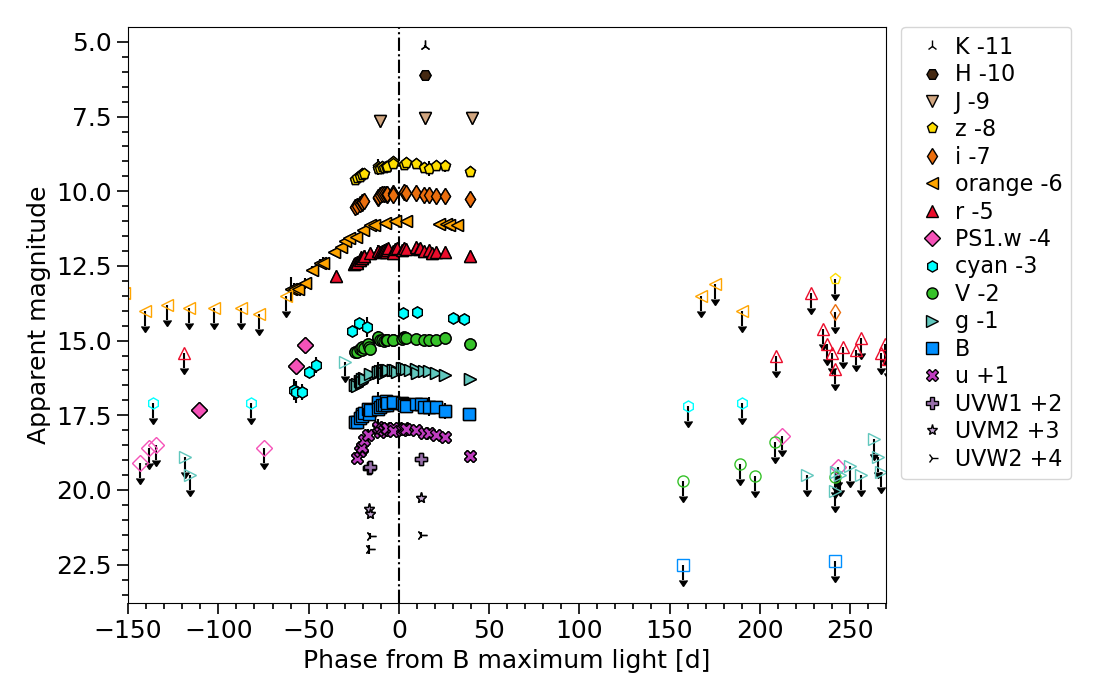} \hfill 
\includegraphics[width=0.7\textwidth]{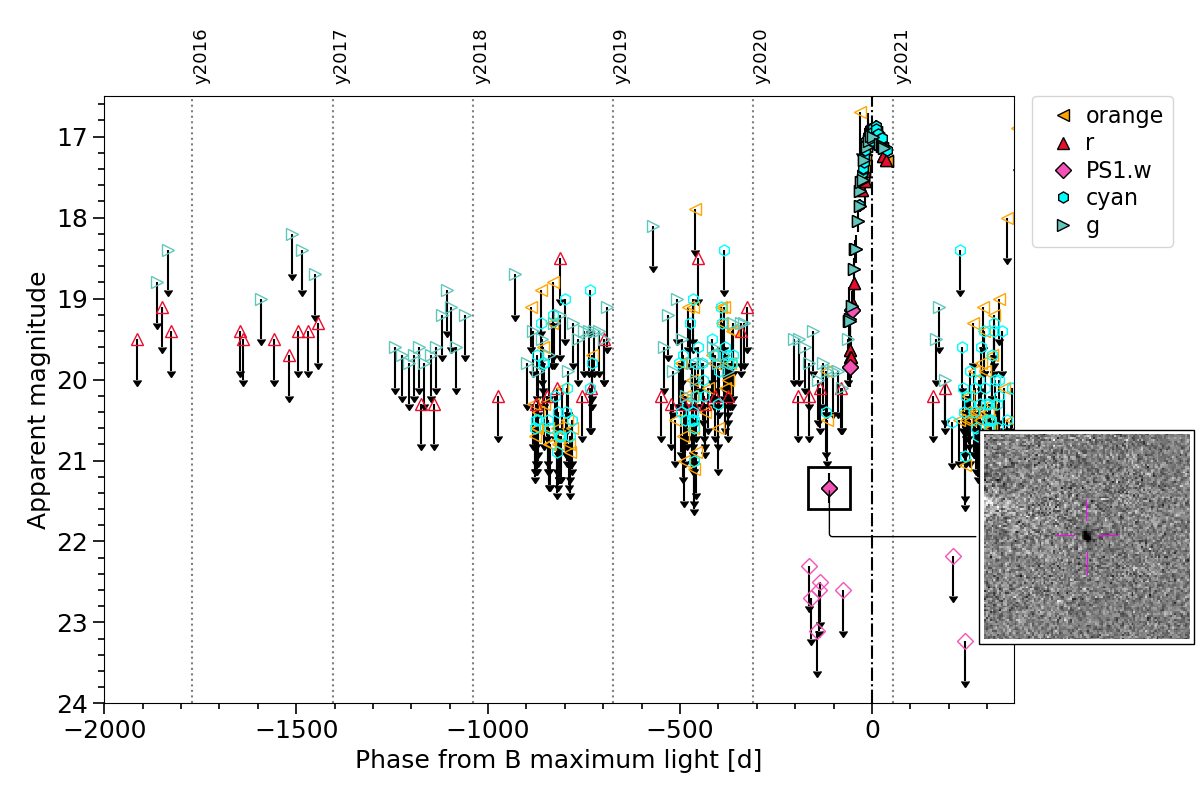} 
\caption{{\it Top:} UV-optical-NIR light curves of \pvb. Upper limits are indicated by empty symbols with arrows. The light curves have been shifted for clarity by the amounts indicated in the legend. {\it Bottom:} Historical light curves of \pvb. Upper limits are indicated by empty symbols with arrows. The light curves have been shifted for clarity by the amounts indicated in the legend. The right insert is a 27$\arcsec \times$ 27$\arcsec$ zoomed in of the transient position in the difference image (observation - reference image) taken on 2020 July 18 in the PS1 $w$-band. The precursor outburst is detected at greater than 5-sigma significance.}
\label{fig_lc}%
\end{figure*}

UV-optical-NIR light curves of \pvb are shown in Fig.~\ref{fig_lc}. The light curves of the SN in all bands are characterised by a plateau-like phase at maximum light. This is reminiscent of the so-called SNe Type IIn-P \citep{mauerhan13b}. 

The optical light curves peak around 60 days after the estimated explosion date (this is best seen in the cyan and orange ATLAS curves), reaching a $B$-band maximum light on MJD = 59159.18 $\pm$ 0.50. The long-lasting rise to the photometric peak is another characteristic of \pvb, in common with the SNe IIn-P. \citet{moriya14} proposed that the slow-rising light curve in interacting SNe results from the interaction with a dense CSM located at a large radius (see Sect. \ref{SNprog}). We also note that the \pvb rise to the maximum light appears to be sharper in the bluer bands \citep[also noted by][for SN~2011ht]{roming12}. The $u$-band light curve in the top panel of Fig. \ref{fig_lc} and Fig. \ref{fig_bluebands} increases by $\sim$ 1 mag from the time of the first observation until peak (in contrast to the $\sim$ 0.6 mag seen in the $B$-band light curve during the same time), and reaches the maximum $\sim$ 6 days earlier than that in the $B$-band. Unfortunately, the NIR follow-up was sparse, and we cannot verify whether the rise is longer in these bands as reported by \citet{mauerhan13b} for SN~2011ht. We notice that late-NIR emission ($>$400 d) was also detected in the case of the Type IIn-P SN~2009kn, probably due to early dust formation between 200-400 d \citep{kankare12}. However, we cannot confirm that this is a typical behaviour of Type IIn-P SNe, as no NIR data are available for other members of this subclass. From our dataset, the \pvb plateau seems to be at least 40 days long from the SN maximum light. However, this is a lower limit because the transient went behind the Sun before the light curve dropped off. Supernovae IIn-P, such as SN~1994W, are expected to decline very rapidly after the plateau phase \citep[see, e.g.][]{chugai04,smith13,chugai16} and follow a decline consistent with that of the radioactive $^{56}$Co decay. In fact, after the seasonal gap, the SN was below the detection threshold ($>$ 22.9 mag in $r$-band).

The \pvb site was observed for several years before the SN discovery (Tables \ref{table_PSph}, \ref{table_ATLASph} and bottom panel of Fig. \ref{fig_lc}). We did not detect a source at the event position brighter than the absolute magnitude -14.5 mag in the ATLAS bands or -12 mag in the PS1 $w$-band, except for the single detection at an absolute magnitude of $M_{\rm PS1 w} \approx -13.8 \pm 0.5$ mag on 2020 July 18 (the Pan-STARRS discovery, $\sim$ 111~d before the $B$-band maximum light and $\sim$ 50 days before the estimated explosion date), that we label as a precursor outburst. This precursor is detected on four separate 45-second images on this night, with no sign of motion on each of them, making the possibility of it being an asteroid unlikely. This source was not seen by the ATLAS survey in either the previous survey's non-detection five days before the outburst or in the monthly stacked image from April to August 2020 down to a limiting magnitude of $M_o \sim$ -15 mag.

\begin{figure}[!ht]
\centering
\includegraphics[width=\columnwidth]{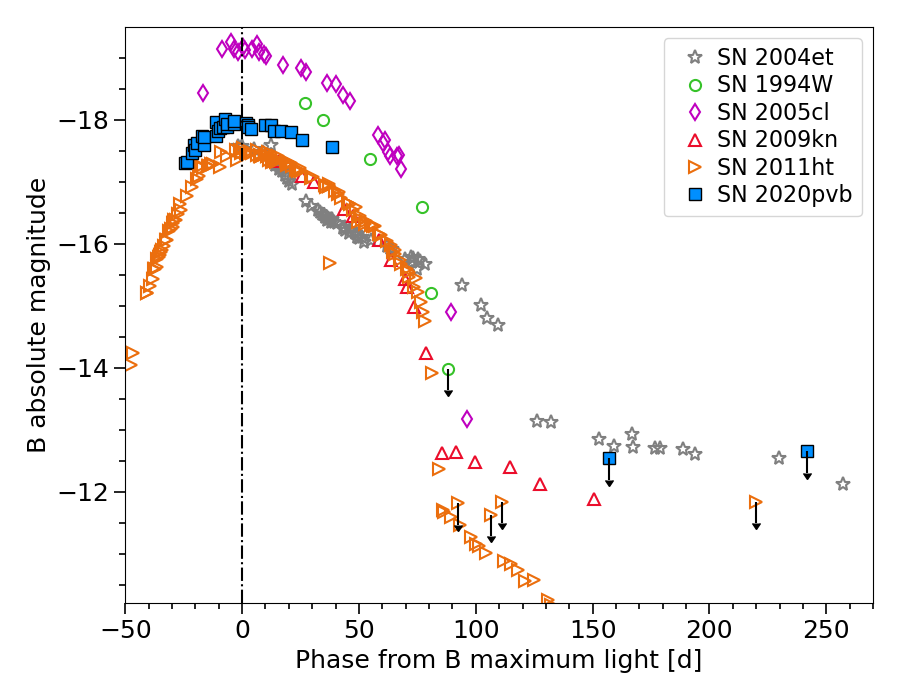}\hfill 
\caption{Absolute $B$ light curve of \pvb, shown together with those of SNe~1994W, 2005cl, 2009kn, 2011ht and the Type~IIP~SN~2004et. Upper limits are indicated by vertical arrows. The dot-dashed vertical line indicates the $B$-band maximum light. The uncertainties for most data points are smaller than the plotted symbols. }
\label{fig_am}%
\end{figure}

Figure \ref{fig_am} shows a comparison between the evolution of the absolute $B$ magnitude of \pvb and those of the \mbox{SNe~IIn-P}~1994W, 2005cl, 2009kn, 2011ht, and the ordinary SN~IIP~2004et \citep{maguire10} that has a similar luminosity to the general IIn-P population. 
The comparison SNe have been corrected for extinction using published estimates, assuming the \citet{cardelli89} extinction law and their respective kinematic distances were scaled assuming H$_0$ = 73.2 $\pm$ 1.3 km\,s$^{-1}$\,Mpc$^{-1}$. \pvb exhibits a broad light curve, similar to those of the SN~1994W-like SNe. This plateau, and the following decline, are different from that of a typical Type IIP such as SN~2004et (Fig. \ref{fig_am}). The absolute $B$ magnitude at the maximum of \pvb is -17.95 $\pm$ 0.30 mag, that is between SN~1994W and SN~2011ht and approximately 1.2 mag fainter than SN~2005cl.

\begin{figure}[!ht]
\centering
\includegraphics[width=\columnwidth]{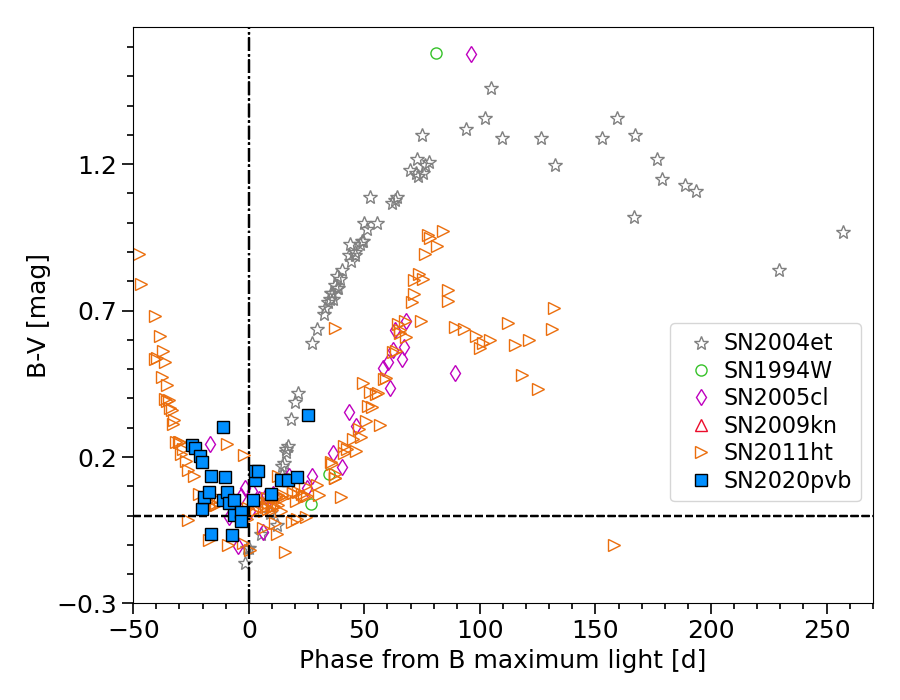} \hfill 
\caption{Intrinsic colour evolution of \pvb, compared with those of SNe~1994W, 2005cl, 2009kn, 2011ht and the Type~IIP~SN~2004et. The dot-dashed vertical line indicates the $B$-band maximum light of \pvb.
} 
\label{fig_colourevol}%
\end{figure}

The ($B-V$)$_0$ colour curve of \pvb (see Fig. \ref{fig_colourevol}) seems to evolve from red to blue at early times, followed by a flattening during the plateau phase and then gradually reddens. By the end of the plateau, \pvb shows a similar colour trend to SNe~1994W, {2005cl,} 2009kn, 2011ht, and yet bluer than SN~2004et. The small evolution in the ($B-V$)$_0$ reflects the SED temperature, which appears to be constant (see Sect. \ref{SNspec}).

\begin{figure}[!ht]
\centering
\includegraphics[width=\columnwidth]{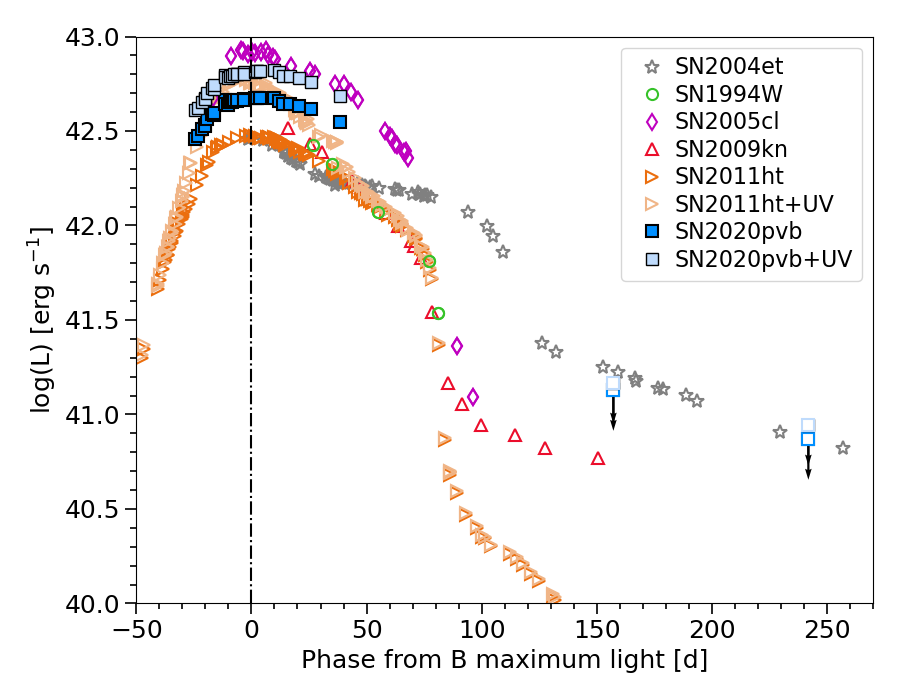} \hfill 
\caption{Pseudo-bolometric light curve of \pvb obtained by integrating optical and NIR bands, compared with those of SNe~1994W, 2005cl, 2009kn, 2011ht and the Type IIP SN~2004et. The UV-optical-NIR pseudo-bolometric light curve of \pvb is also included. The dot-dashed vertical line indicates the $B$-band maximum light of \pvb.} 
\label{fig_bol}%
\end{figure}

We computed the pseudo-bolometric light curve of \pvb, and the comparison SNe, by integrating the flux from the extinction-corrected optical-NIR magnitudes. Fluxes were measured at epochs when $B$-band observations were available. When photometric measurements in one band at given epochs were not available, the flux was estimated by interpolating magnitudes from epochs close in time or, when necessary, by extrapolating the missing photometry assuming a constant colour. We estimated the pseudo-bolometric flux at each epoch by integrating the SED using the trapezoidal rule and assuming zero flux outside the integration boundaries. Finally, the effective fluxes were converted to luminosities using the adopted distance to the SN (see Sect. \ref{SNhostgx}). The bolometric luminosity errors include the uncertainties in the distance estimate, the extinction, and the apparent magnitudes. 
As shown in Fig. \ref{fig_bol}, the pseudo-bolometric plateau of \pvb is similar in shape to SN~1994W-like SNe, although more luminous, except for SN~2005cl. 
Fitting low-order polynomials to the light curve, we estimate a peak luminosity of 4.8 $\pm$ 0.6 $\times$ 10$^{42}$ erg s$^{-1}$. We also include in the figure the pseudo-bolometric light curve of \pvb, including the UV data. This wavelength range contributes $\sim$ 25$\%$ to the SN luminosity. Note, however, that UV photometry is available only at two epochs near maximum light

\subsection{Spectral evolution}\label{SNspec}

\begin{figure*}[!ht]
\centering
\includegraphics[width=\textwidth]{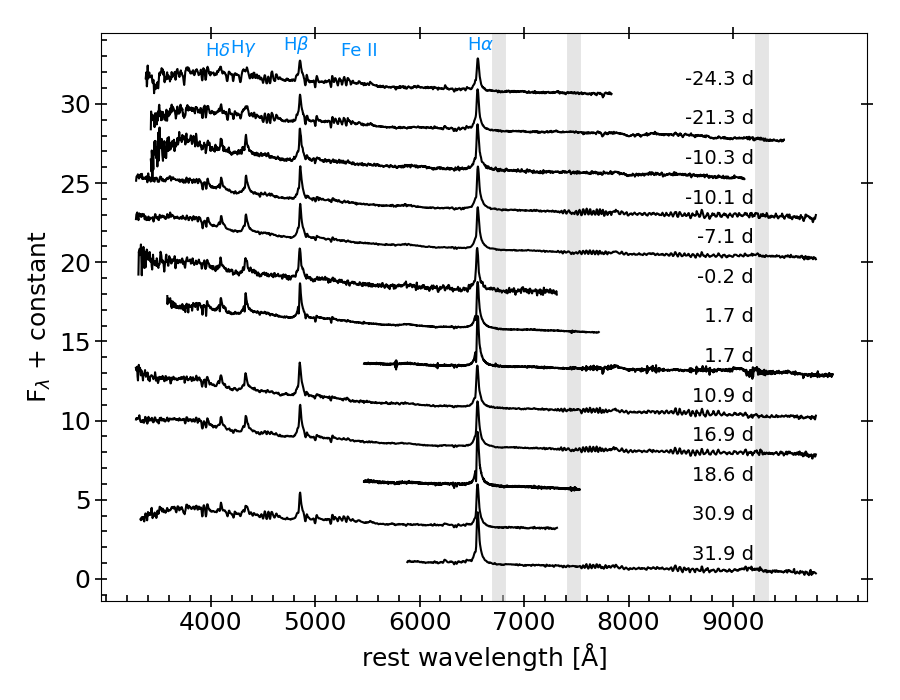} 
\caption{Spectral sequence of \pvb extending from -24.3~d to 31.9~d from maximum light. All spectra have been corrected by redshift. The shaded wavelength regions indicate areas of strong telluric absorption, which has been removed when possible. The locations of the most prominent spectral features are also indicated.}
\label{fig_specevol}%
\end{figure*}

\begin{figure*}[!ht]
\centering
\includegraphics[width=0.7\textwidth]{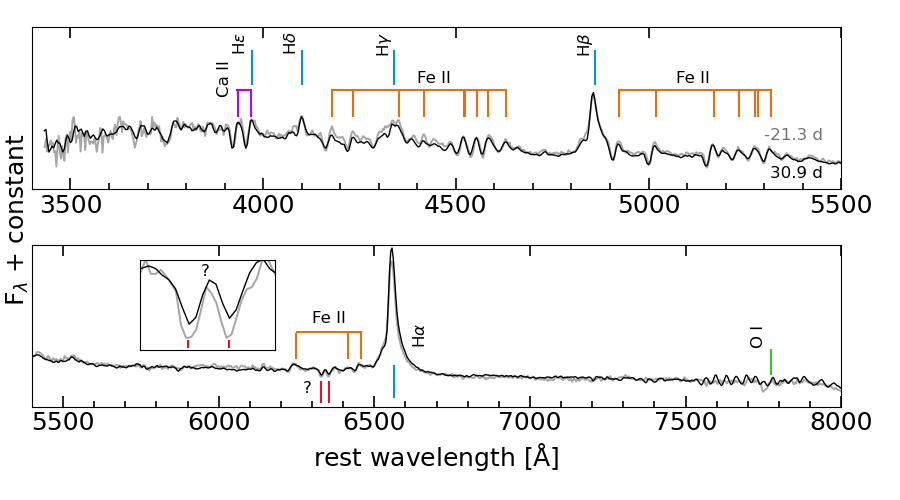} 
\caption{Superposition of the 2020 October 15.88 UTC (-21.3 d, solid grey line) and 2020 December 08.03 UTC (31.9 d, dotted black line) spectra of \pvb. The insert is a zoomed-in of the absorptions
between 6300 and 6400 \AA. The most prominent spectral features are indicated.}
\label{fig_specoverplot}%
\end{figure*}

Figure \ref{fig_specevol} shows the spectral evolution of \pvb, and in Fig.~\ref{fig_specoverplot}, we superpose the second spectrum (taken during the rise time at phase -21.3 d) and the last one (taken at 30.9 d, during the decline) spectra of \pvb.

The spectra of \pvb exhibit a blue continuum with very little evolution (note the similarity in features and line velocities among the spectra of Fig. \ref{fig_specoverplot}). They are dominated by multi-component Balmer lines in emission and many \ion{Fe}{ii} features with narrow P Cygni profiles. \ion{Ca}{ii} H\&K $\lambda\lambda$3933, 3968 features are visible in the first spectrum and throughout the plateau. In the red part of the spectra, it is noteworthy the complete absence of the \ion{Ca}{ii} NIR triplet, features almost always seen in both IIn and IIP SNe. Only a weak feature, possibly \ion{O}{i} $\lambda$7774, is visible. We also notice two small non-identified absorptions between 6300 and 6400 \AA \xspace\xspace present in all spectra. We searched for typical SN lines, e.g. \ion{Fe}{ii} and/or \ion{Sc}{ii} features, but we did not find any convincing identification. One possible identification is with \ion{Si}{ii} $\lambda\lambda$ 6347, 6371 also present in SN~1994W spectra from 21 to 89 d after its explosion \citep{chugai04}.
Alternatively, we also considered the possibility that these absorptions were diffuse interstellar bands (DIBs), although they do not coincide with any of the DIBs listed in \citet{herbig95} or \citet{fan19}. Moreover, DIB intensity is correlated with the column density of \ion{Na}{i} in the line of sight \citep{herbig93}, which we consider negligible in this case. In fact, we do not identify any of the most intense DIBs such as at $\lambda$4428, $\lambda$5780 or $\lambda$6284, also seen in the bright SN~1987A \citep{vladilo87} or extinguished SNe such as SN~2003cg \citep{eliasRosa06}.

\begin{figure*}[!ht]
\centering
\includegraphics[width=\textwidth]{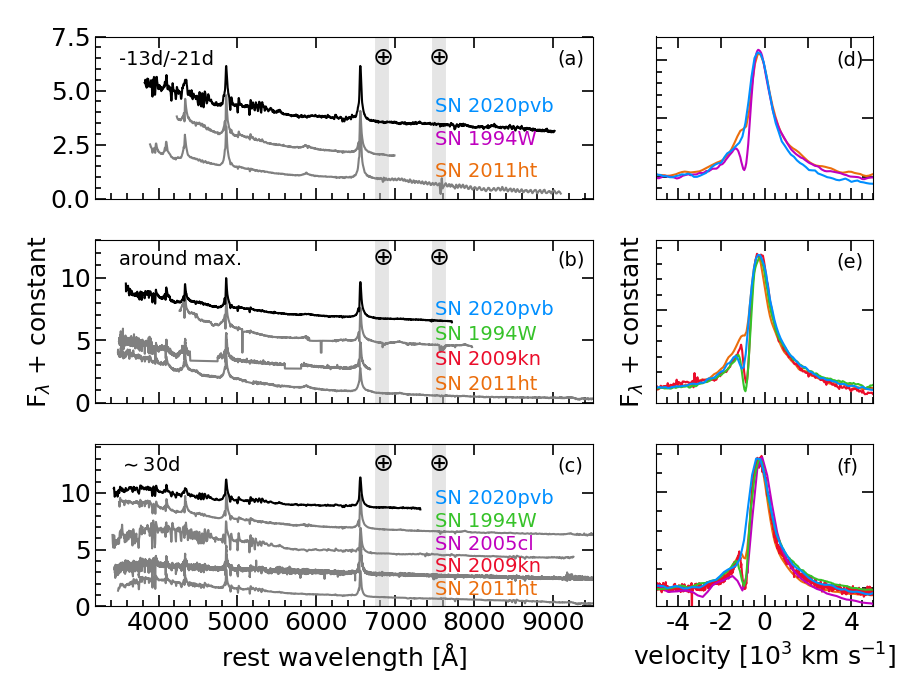} 
\caption{Comparison of \pvb spectra before {\it (a)}, around {\it (b)}, and after {\it (c)} the maximum peak. We also include those of SNe~1994W, 2005cl, 2009kn and 2011ht at similar epochs. The H$\alpha$ profiles are enlarged on the right of each panel and shifted to the peak ({\it (d), (e)} and (f)). The H$\alpha$ profiles of each SN match the colours of their label on the left. All spectra have been corrected for their host-galaxy recessional velocities and extinctions (values adopted from the literature).}
\label{fig_speccomp}%
\end{figure*}

In Fig. \ref{fig_speccomp}, we compare the spectra of \pvb with those of SNe~1994W \citep{chu04}, 2005cl \citep[][WISeREP]{kiewe12}, 2009kn \citep{kankare12} and 2011ht \citep[][Padova-Asiago SN archive\footnote{\url{https://sngroup.oapd.inaf.it}}]{humphreys12} at similar epochs. Following \citet[][for SN~2011ht]{chugai16} and \citet[][for SN~1994W]{dessart16}, these SNe have been argued to result from the interaction of low-mass ejecta with an extended, slow and massive outer shell. Therefore, their spectral continuum (or photosphere) form within the extended shell which is dense, partially ionized and moves with velocities of 500-1000 \kms. All objects share the same characteristics with a blue continuum dominated by Balmer lines. The forest of narrow P Cygni \ion{Fe}{ii} and a possible resulting minimal blue excess that is seen in \pvb before the maximum light is visible also in the other reference SNe, such as SNe~2005cl and 2011ht after the maximum peak. We also highlight the similarity of the H$\alpha$ profiles (see the right side of each panel in Fig. \ref{fig_speccomp}), although with some velocity variations in the blue wings. At early times, \pvb shows a symmetric H$\alpha$ profile, unlike SNe 1994W and 2011ht. As time progresses, the line loses its blue extended wing becoming narrower and displaying a blue shifted absorption line at ~$\sim$ 900 \kms, at a similar velocity to those seen in SNe 1994W, 2009kn and 2011ht. The appearance of the absorption line means that the H above the photosphere is partially recombined. 
At phase $\sim$ 30 d, the absorption component of \pvb is much less prominent compared to the other SNe, which could be due to a poorer spectral resolution in the \pvb data or could indicate that the optical depth of the electron scattering is decreasing faster in \pvb. 

\begin{figure}[!ht]
\centering
\includegraphics[width=\columnwidth]{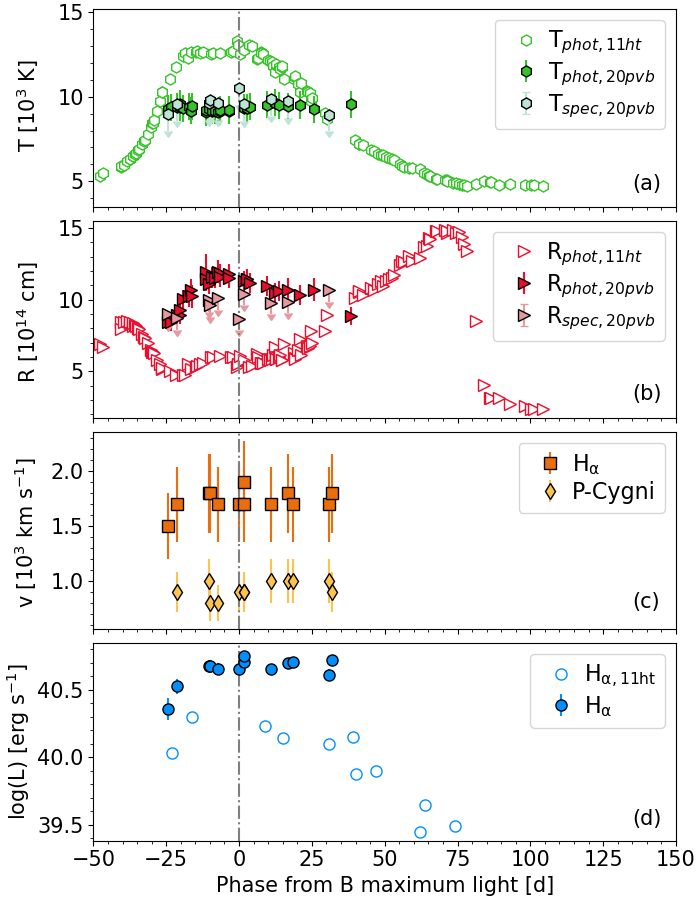} 
\caption{Evolution of the best-fit blackbody temperatures {\it (a)}, photospheric radius {\it (b)}, FWHM and blue-shift evolution for the whole profile and P Cygni H$\alpha$ emission {\it (c)}, and evolution of the total luminosity of H$\alpha$ {\it (d)} of \pvb and SN~2011ht. The dot-dashed vertical line indicates the $B$-band maximum light of \pvb.}
\label{fig_tempvelevol}%
\end{figure}

We estimate the photospheric temperatures by fitting with a blackbody function the spectral continuum and the SED obtained with broadband photometry (see \citealt{pastorello21} for a detailed description of the procedure). We consider them upper limits due to the possibility of a blue excess at wavelengths shorter than 5500A (please see the previous paragraph). As shown in Fig. \ref{fig_tempvelevol} (panel a), there is a small scatter in temperature during the \pvb temporal evolution, ranging from 9000 to 10500 K. These temperatures are higher than the recombination temperature of H. SN~2011ht is also significantly hotter and shows a constant temperature during the plateau phase, but there is clear evolution before and after it. On the contrary, the blackbody radius of the photosphere estimated through the Stefan–Boltzmann law exhibits a slow evolution peaking at 12 $\times$ 10$^{14}$ cm at $\sim$ 11d before the $B$-band maximum light, and then decreases to about 8.4 $\times$ 10$^{14}$ cm at 38.6 d (panel b of Fig. \ref{fig_tempvelevol}). 
Note that these radii estimates are approximate, owing to the assumptions made to derive the temperatures and the luminosities of \pvb. In particular, the derived temperatures are more likely lower-limit estimates since we have assumed blackbody spectra without taking into account effects such as the metal line blanketing and emission/absorption features.\\

\begin{figure}[!ht]
\centering
\includegraphics[width=\columnwidth]{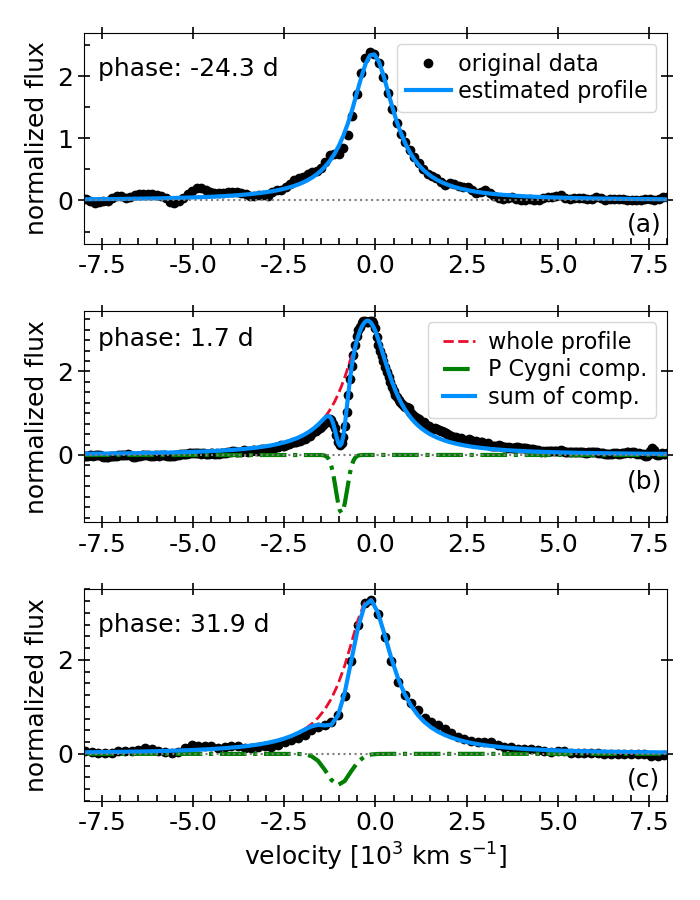} 
\caption{ Deblend of the H$\alpha$ emission line of \pvb at -24.3, 1.7 and 31.9 d from the $B$-band maximum light.}
\label{fig_specHalphadecomp}%
\end{figure}

The Balmer line profiles, particularly those of H$\alpha$, seem to consist of multiple components. We decomposed the H$\alpha$ line profile at all epochs using a least-squares minimisation multi-component fit. Figure \ref{fig_specHalphadecomp} presents the results of this fit at some representative epochs: rise (-24.3 d), around (1.7 d), and after (31.9 d), the $B$-band maximum. The profiles are reproduced using a single Lorentzian profile for all the spectra, including an additional Gaussian component in absorption when P Cygni features were visible (at phases $>$ -21.3 d). The velocity estimates for the emission components (panel c of Fig. \ref{fig_tempvelevol}) are derived by measuring their FWHM, while those of the absorbing gas are estimated from the wavelengths of the P Cygni minima. The FWHM of the H$\alpha$ emission (after correction for instrumental resolution\footnote{We first corrected the measured FWHM for the spectral resolution ($width = \sqrt{FWHM^{2} - res^{2}}$) and then computed the velocity (\mbox{$v = (width/\lambda_0) \times c$)}.}) remains nearly constant, with an average value of $\sim$1700 \kms, while the absorption feature in the blue wing of H$\alpha$ (shallow at late times) has a constant blue-shift of $\sim$ 900 \kms.

The observed H$\alpha$ emission peak is blue-shifted by about 100 \kms regarding the rest wavelength with a red tail extending to a higher velocity than the blue wing (see Fig. \ref{fig_specHalphadecomp}). This behaviour has been seen in other SN~1994W-like events and is explained by \citet{dessart05} as the results of multiple electron scattering for photons trapped in an optically thick emitting region.

The integrated luminosity of H$\alpha$ (panel d of Fig. \ref{fig_tempvelevol}) evolves similarly as the broad band light curves. It shows a rise from $L_{\rm H\alpha}$(-24.3d) = 2.3 $\pm$ 0.4 $\times$ 10$^{40}$ erg\,s$^{-1}$ then remains roughly constant (since $\sim$ 10 days before the $B$-band peak) at $\sim$ 4.8 $\times$ 10$^{40}$ erg\,s$^{-1}$ during the plateau phase. \pvb has a substantially higher H$\alpha$ luminosity than SN~2011ht (by a factor of $\sim$ 2) but it is fainter than SNe~1994W (see, e.g. Fig. 8 of \citealt{mauerhan13b}). \pvb also shows an approximately constant luminosity evolution of H$\alpha$ during the plateau phase. Similarly to SN 2011ht, we could expect a later decline. Unfortunately, our transient went behind the Sun, and we could not take any other measures. 

\section{Identification and nature of the progenitor candidate}\label{SNprog}

We localise the \pvb position on the ACS/WFC images by performing differential astrometry between these and a $V$-band NOT+ALFOSC image taken on 2020 November 08, with a seeing of 0$\farcs$8. We find \pvb position to correspond to the pixel coordinates 2389.80, 3372.46 on the F606W image with an associated RMS uncertainty $\sim$ 0$\farcs$11 ($\approx$ 2.2 pixels).

\begin{figure*}[!ht]
\centering
\includegraphics[width=0.7\textwidth]{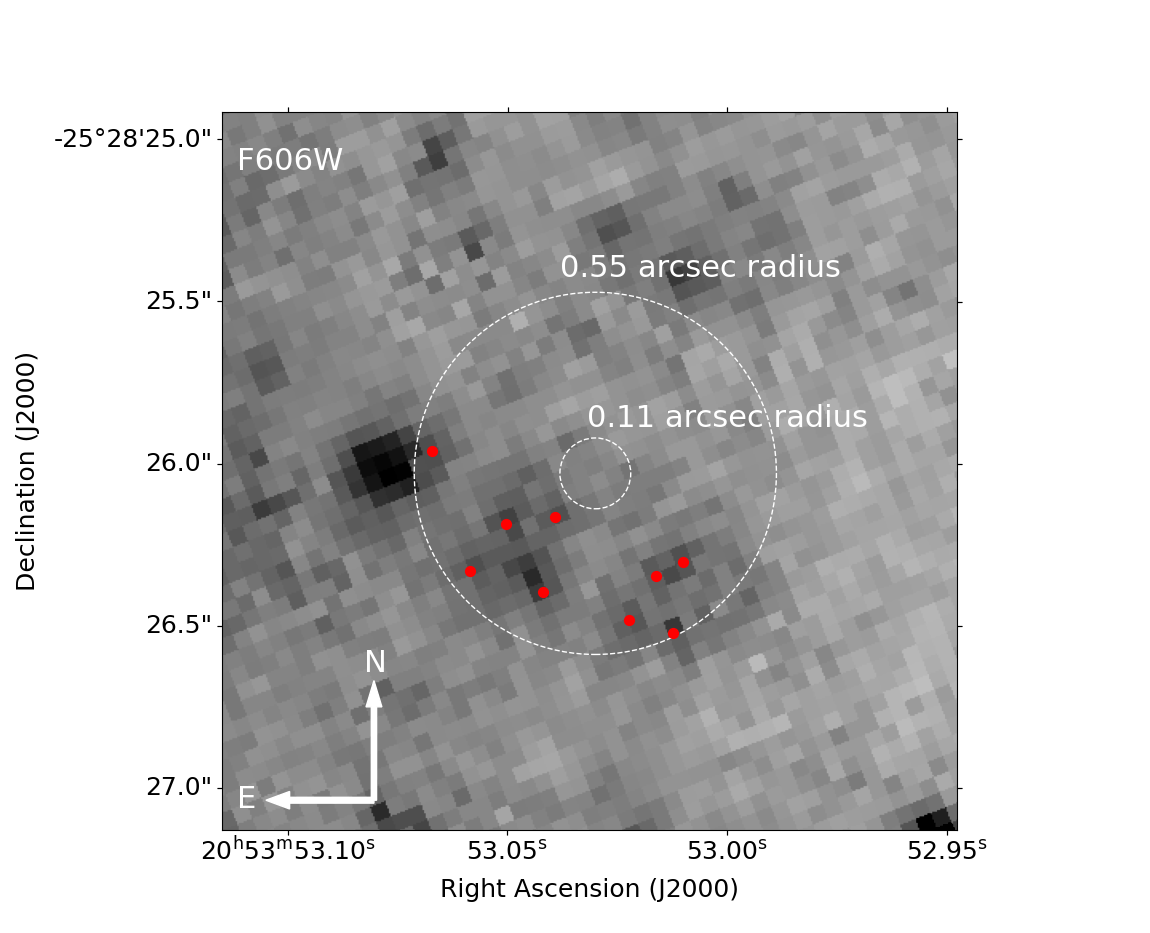} 
\caption{2$\farcs$3 $\times$ 2$\farcs$3 cutouts from the 2017 HST+ACS/WFC image of the \pvb site. The dashed circles are centred at the \pvb position and indicate a 0$\farcs$11 and 0$\farcs$55 (5$\sigma$ significance) positional uncertainties based on the RMS uncertainty obtained from the astrometry. Red points are the likely stellar sources detected by {\sc dolphot} at these areas.} 
\label{fig_progcand}%
\end{figure*}

We measured the brightness of the progenitor candidates using the {\sc dolphot} photometry package. We did not detect any source within an area of radius equal to the RMS uncertainty of the \pvb position. Instead, we identified several sources flagged as `object type = 1', meaning they are likely stellar, at 5$\sigma$ positional confidence within an area of radius 0$\farcs$55. They are shown in Fig. \ref{fig_progcand}. These sources are consistent with the position of \pvb, and therefore, the candidate progenitor of this SN could be one of them with a mag (in the Vega system) between 26.2 and 26.9 mag in F606W (corresponding to the brightest and faintest magnitude sources among those detected) or otherwise fainter than F606W = 26.9 mag.

\begin{figure}[!ht]
\centering
\includegraphics[width=\columnwidth]{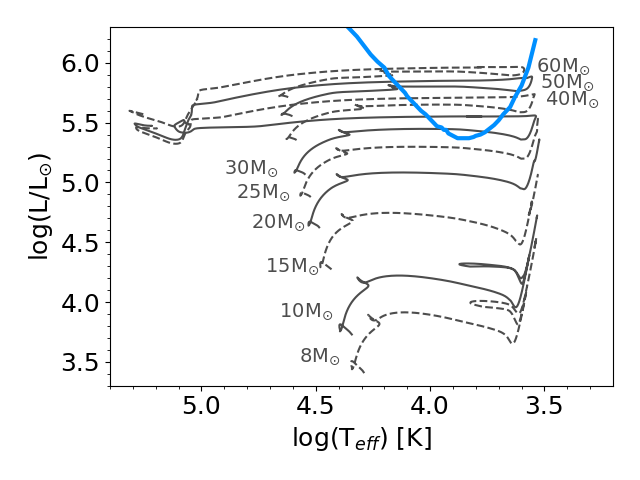} 
\caption{Hertzsprung-Russell (HR) diagram showing the \pvb bolometric luminosity upper limit as a function of the effective temperature (solid blue line). The solid and dashed grey lines show single star evolutionary tracks from 8 to 60 $M_\sun$ from the single star BPASS models \citep[v2.2.1;][]{Eldridge2017,stanway18}, assuming solar metallicity.}
\label{fig_hrdiag}%
\end{figure}

Correcting for the total extinction and distance assumed for \pvb (see Sect. \ref{SNhostgx}), we find that the absolute magnitude of the progenitor was $M_{\rm F606W} \gtrsim$ -8.7 mag. Not having colour information at the HST epoch, the progenitor initial mass is not well constrained and we can only set an upper limit of $M_{\rm ini}$ $\lesssim$ 50 $M_\sun$, depending on the bolometric correction (see Fig. \ref{fig_hrdiag}). Therefore, the \pvb progenitor can be a low-mass luminous blue variable (LBV; these stars are usually suggested to be the progenitors of Type~IIn~SNe -- \citealt{galyam07,smartt15}) or a less massive star. 
 
\section{Discussion and summary}
\label{SNdiscussion}

\subsection{Nature of \pvb}
\pvb is a relatively bright ($M_B$ = -17.95 $\pm$ 0.30 mag) Type IIn SN with a plateau phase, followed likely by a rapid decline. Persistent spectral signatures of interaction and a plateau-like photometric evolution classify \pvb as a Type IIn-P SN \citep{mauerhan13b} akin to SNe~1994W, 2009kn and 2011ht. These transients share relatively high luminosity, are associated with low kinetic energy, and have a low luminosity during the tail, implying a low $^{56}$Ni mass \citep[e.g.][]{chugai16}. Following the above, we could consider a similar explosion scenario for \pvb requiring a low $^{56}$Ni mass. We roughly estimate a $^{56}$Ni mass $\leq 0.1 M_{\sun}$ from the first upper limit obtained after the solar conjunction (phase 157 d) and using the formula given by \citet{hamuy03}. Note that a $^{56}$Ni mass of 0.023 $M_{\sun}$ was estimated for SN~2009kn \citep{kankare15}, the SN IIn-P with the most luminous light curve tail (see top panel of Fig. \ref{fig_am}).

The observational characteristics of \pvb are so similar to those of SNe~1994W and 2011ht that one could postulate that these transients have a similar physical interpretation. Figure \ref{fig_bol} shows an abrupt luminosity decay after the \pvb plateau phase, with a late time flux upper limit of 1.5 $\times$ 10$^{41}$ erg~s$^{-1}$ at phase 157 d and 8.8 $\times$ 10$^{40}$ erg~s$^{-1}$ at phase 242 d. This could indicate that the $^{56}$Ni mass of \pvb is lower than 0.015 $M_{\sun}$, upper $^{56}$Ni mass limit estimated for SN~1994W \citep{sollerman98}. 

SN~1994W-like SNe display similar strong Balmer lines with multiple components and a forest of narrow P Cygni lines of \ion{Fe}{ii} (see, e.g. Fig. \ref{fig_speccomp}). \pvb shows shallow absorption features in the blue wing of H$\alpha$, H$\beta$ and \ion{Fe}{ii} $\lambda$5018 with an average blue-shift of $\sim$800-950 \kms. These lines could arise from an outer, expanding shell lost by the progenitor star \citep[e.g. SNe 1994aj and 1996L;][respectively]{Benetti1998,Benetti1999}. 
Considering the shell velocity as derived from the minima of the absorption (900 \kms) and the inner shell radius close to that of the photosphere (8.5 $\times$ 10$^{14}$ cm; we have taken the first value of radius derived at phase -24.4 d), we estimate that the shell surrounding \pvb has\footnote{We approximate the shell density, in terms of the hydrogen concentration for a normal abundance, as $n \gg 3\times10^8 v_{3} r_{15}^{-1}cm^{-3}$, where $v_3$ is the shell velocity in units of 10$^3$ \kms, and $r_{15}$ is the shell radius in units of 10$^15$ cm.} an $n \gg$ 10$^8$ cm$^{-3}$. However, the velocities measured in these absorption minima might not necessarily be associated to a steady wind in the outer unshocked shell, but in the interaction region between a faster inner shell and an outer slower shell. This scenario was presented by \citet{dessart16} and applied to SN~1994W. As described in this model, the fastest material of the inner shell, with some contribution of the accelerated outer shell material, is piled up in this region between the two shells, moving outwards at a constant velocity of about 900 km/s.

Another spectral characteristic of \pvb is its nearly flat Balmer decrement of H$\alpha$/H$\beta$ $\sim$ 2, suggesting the contribution of radiative transitions or collisional thermalization \citep{chugai04}.\\

\subsection{Progenitor scenario}
As for other similar transients, it is challenging to identify a unique progenitor scenario for \pvb. Here we discuss two simple possibilities:

\noindent{Option A.-} Massive progenitor: Through HST images, we have estimated that the \pvb progenitor has a luminosity of $\log(L/L_{\sun}) \lesssim$ 5.4, which is consistent with a single star candidate with an initial mass $M_{ini}$ $\lesssim$ 50 $M_\sun$. 
Recently, \citet{matsumoto22} have modelled the optical precursor outburst detected from a sample of core-collapse SNe considering two scenarios: a single eruption or a continuous wind. Using their equation 30, we can estimate the wind mass-loss rate ($\dot{M}$) in terms of observed precursor properties. With the previously derived progenitor values and considering 900 \kms as the average wind velocity estimated from the narrow H$\alpha$ P Cygni profile of \pvb (see Fig.~\ref{fig_tempvelevol}), we obtain $\dot{M}$ = $6 \times 10^{-5}$ $M_{\sun}$ yr$^{-1}$. This value is consistent with that of hypergiant stars or LBV winds, particularly in outburst \citep[e.g.][]{debeck10,smith17}. 
Within this massive progenitor option, an even valid scenario could be that described by \citet{heger03} where a progenitor of around 40 $M_\sun$ could generate a weak Type~II~SN. In this case, the ejecta interacts with the possible mass lost by the star just before the explosion (for example, during the detected outburst). Then, once the interaction and recombination of the ejecta finishes, it cools down quickly because most of the $^{56}$Ni is swallowed by the black hole.\\

\noindent{Option B.-} Moderate/low mass progenitor: The \pvb observables are also consistent with a low mass progenitor surrounded by a dense CSM scenario \citep{sollerman98,kankare12,mauerhan13b,chugai16}, although it is unclear what could be the mechanism behind a precursor eruption from an 8-10 $M_\sun$.\\

As discussed in Sect. \ref{SNprog}, PS1 detected an outburst at PS1 $w$-band $\sim$ -13.8 mag around 111 days before the $B$-band maximum light ($\sim$ 50 days before the estimated explosion date). Interestingly, \citet{fraser13b} reported a pre-SN eruption also for SN~2011ht $\sim$6 months before its explosion at $M_{\rm z}$ = -11.8 mag. These events appear to confirm the \citet{ofek14b} and \citet{strotjohann21} findings that the outbursts of interacting SNe are frequent in the last few months before the SN explosions. We estimate a $L_{\rm pre} \approx$ 9.6 $\times$ 10$^{40}$ erg s$^{-1}$ for the pre-SN outburst of \pvb. By directly multiplying this luminosity by its maximum possible duration, $t_{\rm pre}$ = 48 d\footnote{Since the \pvb's precursor was detected on just a single day, and there are Pan-STARRS survey non-detections both 24 days before and after the outburst, we assumed $t_{\rm pre}$ = 48 d and assessed it as an upper limit. We want to note that the ATLAS survey had non-detections only five days before the outburst and also three days later. This would make a $t_{\rm pre}$ of about eight days, reducing the radiated energy and the other estimated outburst parameters by about 17$\%$. However, since the ATLAS survey's non-detections are shallower than those obtained from the Pan-STARRS survey, we still consider the limit estimations from $t_{\rm pre}$ = 48 d to be more robust.}, the radiated energy of the outburst was $<$~4~$\times$~10$^{47}$~erg, consistent with that observed for SN~2011ht \citep{fraser13b}\footnote{See also \citet{pastorello19} who argue that the outburst of SN~2011ht was the outcome of a merger based on the similarities of its shape and duration of the light curve with that of luminous red novae.}. On the other hand, taking as reference the gas velocity measured from the \ion{Fe}{ii} $\lambda$5018 line (that is free from blending), its broad component (FWHM $\sim$ 800 \kms) may be connected with the dense thin shell at the boundary between the SN and the CS gas \citep{chugai01,chugai03}. Therefore, assuming again 8.5 $\times$ 10$^{14}$ cm as the shell radius, we estimate that this shell was ejected about four months before the explosion. This finding is roughly consistent with the date of the precursor outburst detected by the Pan-STARRS survey at the \pvb position.

Using the \citet{matsumoto22} equations 25, 27, and 28 for a single eruption scenario based on the `Popov formulae' \citep{popov93}, we can estimate an upper limit for the total ejected mass $M_{\rm shell}$, the outer CSM radius $R_{\rm shell}$ and the density profile of the precursor outburst $\rho_{\rm shell}$. Assuming $v_{\rm ej}$ (velocity of the outburst ejecta) of 900 \kms (shell velocity derived from the minima of the H$\alpha$ absorption), $L_{\rm pl}$ = $L_{\rm pre} \approx$ 9.6 $\times$ 10$^{40}$ erg s$^{-1}$ and $t_{\rm pl}$ = $t_{\rm pre}$ = 48 d as precursor luminosity and total duration time before the SN explosion, respectively, we obtain $M_{\rm shell}$ $\leq$ 0.3 $M_{\sun}$, $R_{\rm shell}$ $>$ 4 $\times$ 10$^{14}$ cm and $\rho_{\rm shell}$ $<$ 2.5 $\times$ 10$^{-12}$ g cm$^{-3}$.\\

Both \cite{chugai04} and \cite{dessart09} modelled the observables of SN~1994W, the prototype of this transient family, finding a need for a contribution from the interaction of the SN ejecta with a dense expanded CS envelope. Another prototypical object of this class is SN~2011ht, which is observationally similar to SN 2020pvb, although very early spectra are not available for the latter. The earliest spectrum of SN 2011ht showed a continuum temperature of $\sim$ 7000K and narrow absorption lines \citep[see Fig. 17 in][]{pastorello19}. To explain the unusual early spectrum and the slow photometric evolution, \citet{roming12} suggested that SN~2011ht exploded in a discontinuous CSM, with the bulk of the CSM located at a relatively large distance. In this scenario, the ejecta-CSM interaction would start with some 
delay. Given the similarity between the two SNe, we cannot rule out a similar scenario for \pvb. Some time later, \cite{chugai16} claimed that the light curves and the low expansion velocity of SN~2011ht, are consistent with a low energy explosion ($< 10^{50}$ erg) and small ejecta mass ($\lesssim$ 2 $M_{\sun}$) interacting with a CS envelope of 6-8 $M_{\sun}$ and radius of $\sim$ 2 $\times 10^{14}$ cm. The flat light curves suggest that the CS interaction started soon after the explosion, where the CSM had to be ejected shortly before the core collapse. In the same context of a moderate/low mass progenitor, \citet{chunhui22} elaborate a somewhat different scenario as an explanation to the type IIn-P subclass calling for a 10 $M_{\sun}$ RSG progenitor which had a precursor outburst of $\sim$ 10$^{46}$ erg some months before it explodes (with final energy of the order of 10$^{51}$ erg). Single outbursts could be caused by a temporarily concentrated injection of energy, probably due to dynamical instabilities of nuclear-burning origin deep within the star, which generates a shock wave that propagates radially outwards and detaches a part of the stellar envelope \citep[e.g.][]{dessart10,kuriyama20,matsumoto22}. They can also be responsible for outburst ejecta masses $\lesssim$ 0.27 $M_{\sun}$ and v$_{ej}$ of 900 \kms \citep{matsumoto22}. Another possibility behind such violent ejections is explosive flashes of degenerate Neon in $<$ 8 $M_{\sun}$ progenitors that are expected months to a few years before the SN explosion and can eject part of the hydrogen envelope with velocities of a few 100 \kms \citep{woosley02,fraser13}. However, such flashes have not been reproduced by more recent models \citep{umeda12,chugai16}. 

Several authors also suggest an alternative scenario, that is a low-energy electron capture explosion (ECSN) of a super-AGB star with strong CSM interaction (e.g. \citealt{mauerhan13b,smith13}; however, see also \citealt{chunhui22}).
\citet{kozyreva21} presented the light curves for an explosion of a super-AGB model with an initial ZAMS mass of 8.8 $M_{\sun}${} which exploded as an ECSN \citep{stockinger20}. The default explosion of this model does not match \pvb. Therefore, we use a modified model scaling up the density of the SN ejecta. The resulting model has an ejecta mass of 0.4~$M_{\sun}${}, which corresponds to the same evolutionary model with a truncated radius of 400~$R_{\sun}${}. As suggested in \cite{kozyreva21}, the truncated model imitates the effect of binarity. We surround the modified model with a wind-like CSM, i.e. with the CSM density $\rho_{CSM}\sim r^{-2}$, and run simulations with the hydrodynamics radiative transfer code {\sc STELLA} \citep{blinnikov93,blinnikov06}. For the best-fit models, the total mass of CSM is 2.2~$M_{\sun}$ and 2.9~$M_{\sun}$, for the e88W1 and e88W2 models, correspondingly, and the CSM radius of $1.5\times 10^{15}$~cm (see Fig.~\ref{fig_ECSNmodels}). The luminosity and the shape of the main peak of the modelled light curves match reasonably well the \pvb's light curves, although the synthetic light curves decline faster after the peak. \\

We stress that we cannot confidently rule out any progenitor scenario presented here. In recent years several interacting supernovae have been discovered with precursor events \citep[e.g.][]{ofek14b,jacobson-galan22}. These transients highlight a gap in our current understanding of the final stages of a star's life, as the conditions responsible for these eruptions are unclear. With continued attention towards these precursor events, and the progenitors themselves, we will better understand how massive stars behave shortly before their death.


\begin{acknowledgements}
      N.E.R. thanks T. Matsumoto for useful discussions.
      We thank the staff at the different observatories for performing the observations.

      N.E.R., S.B., E.C., A.P., A.R., G.V. acknowledges support from the PRIN-INAF 2022, `Shedding light on the nature of gap transients: from the observations to the models'.
      
      S.J.B. would like to thank their support from Science Foundation Ireland and the Royal Society (RS-EA/3471).
      M.F. is supported by a Royal Society - Science Foundation Ireland University Research Fellowship.

      M.G. is supported by the EU Horizon 2020 research and innovation programme under grant agreement No 101004719.

      N.I. was partially supported by Polish NCN DAINA grant No. 2017/27/L/ST9/03221

      S.M. was funded by the Research Council of Finland project 350458.

      S.M. acknowledges support from the Magnus Ehrnrooth Foundation and the Vilho, Yrj\"{o}, and Kalle V\"{a}is\"{a}l\"{a} Foundation.

      T.E.M.B. acknowledges financial support from the Spanish Ministerio de Ciencia e Innovaci\'on (MCIN), the Agencia Estatal de Investigaci\'on (AEI) 10.13039/501100011033 under the PID2020-115253GA-I00 HOSTFLOWS project, from Centro Superior de Investigaciones Cient\'ificas (CSIC) under the PIE project 20215AT016 and the I-LINK 2021 LINKA20409, and the program Unidad de Excelencia Mar\'ia de Maeztu CEX2020-001058-M.

      M.N. is supported by the European Research Council (ERC) under the European Union’s Horizon 2020 research and innovation programme (grant agreement No.~948381) and by a Fellowship from the Alan Turing Institute.

      Support for G.P. is provided by the Ministry of Economy, Development, and Tourism’s Millennium Science Initiative through grant IC120009, awarded to The Millennium Institute of Astrophysics (MAS).

      T.M.R. acknowledges the financial support of the Vilho, Yrj{\"o} and Kalle V{\"a}is{\"a}l{\"a} Foundation of the Finnish academy of Science and Letters. 

      S.J.S. is funded through STFC grant ST/T000198/1

      L.T. acknowledges support from MIUR (PRIN 2017 grant 20179ZF5KS).

      Y.-Z. Cai is supported by the National Natural Science Foundation of China (NSFC, Grant No. 12303054) and the International Centre of Supernovae, Yunnan Key Laboratory (No. 202302AN360001).

      C.P.G. acknowledges financial support from the Secretary of Universities and Research (Government of Catalonia) and by the Horizon 2020 Research and Innovation Programme of the European Union under the Marie Sk\l{}odowska-Curie and the Beatriu de Pin\'os 2021 BP 00168 programme, from the Spanish Ministerio de Ciencia e Innovaci\'on (MCIN) and the Agencia Estatal de Investigaci\'on (AEI) 10.13039/501100011033 under the PID2020-115253GA-I00 HOSTFLOWS project, and the program Unidad de Excelencia Mar\'ia de Maeztu CEX2020-001058-M.
      This work made use of v2.2.1 of the Binary Population and Spectral Synthesis (BPASS) models as last described in \citet{Eldridge2017,stanway18}.

      This work was funded by ANID, Millennium Science Initiative, ICN12\textunderscore009.
      This work makes use of observations from the Las Cumbres Observatory global telescope network.  
      
      This work is based on observations made with the Nordic Optical Telescope, owned in collaboration by the University of Turku and Aarhus University, and operated jointly by Aarhus University, the University of Turku and the University of Oslo, representing Denmark, Finland and Norway, the University of Iceland and Stockholm University at the Observatorio del Roque de los Muchachos, La Palma, Spain, of the Instituto de Astrofisica de Canarias;
      
      ALFOSC, which is provided by the Instituto de Astrofisica de Andalucia (IAA) under a joint agreement with the University of Copenhagen and NOT;
      
      the Gran Telescopio Canarias (GTC), installed in the Spanish Observatorio del Roque de los Muchachos of the Instituto de Astrofísica de Canarias, in the island of La Palma; 
      
      the Liverpool Telescope operated on the island of La Palma by Liverpool John Moores University in the Spanish Observatorio del Roque de los Muchachos of the Instituto de Astrofisica de Canarias with financial support from the UK Science and Technology Facilities Council;
      
      the Italian Telescopio Nazionale Galileo (TNG) operated the island of La Palma by the Fundación Galileo Galilei of the INAF (Istituto Nazionale di Astrofisica) at the Spanish Observatorio del Roque de los Muchachos of the Instituto de Astrofísica de Canarias; 
      
      the STELLA robotic telescopes in Tenerife, an Leibniz-Institute for Astrophysics Potsdam (AIP) facility jointly operated by AIP and Instituto de Astrofísica de Canarias.
       
      Observations from the NOT were obtained through the NUTS2 collaboration which is supported in part by the Instrument Centre for Danish Astrophysics (IDA). We acknowledge funding to support our NOT observations from the Finnish Centre for Astronomy with ESO (FINCA), University of Turku, Finland (Academy of Finland grant nr 306531). 
      
      Based on observations collected at the European Southern Observatory under ESO programmes ID 1103.D-0328 and 106.216C, as part of ePESSTO+ (the advanced Public ESO Spectroscopic Survey for Transient Objects Survey; PI: Inserra).

      LCO data have been obtained via an OPTICON proposals (IDs: SUPA2020B-002 and OPTICON 20B/003). The OPTICON project has received funding from the European Union’s Horizon 2020 research and innovation programme under grant no. 730890

      Pan-STARRS is a project of the Institute for Astronomy of the University of Hawaii, and is supported by the NASA SSO Near Earth Observation Program under grants 80NSSC18K0971, NNX14AM74G, NNX12AR65G, NNX13AQ47G, NNX08AR22G, 80NSSC21K1572 and by the State of Hawaii. The Pan-STARRS data are processed at Queen's University Belfast enabled through the STFC grants ST/P000312/1 and ST/T000198/1.

      This work has made use of data from the Asteroid Terrestrial-impact Last Alert System (ATLAS) project. The Asteroid Terrestrial-impact Last Alert System (ATLAS) project is primarily funded to search for near earth asteroids through NASA grants NN12AR55G, 80NSSC18K0284, and 80NSSC18K1575; byproducts of the NEO search include images and catalogs from the survey area. This work was partially funded by Kepler/K2 grant J1944/80NSSC19K0112 and HST GO-15889, and STFC grants ST/T000198/1 and ST/S006109/1. The ATLAS science products have been made possible through the contributions of the University of Hawaii Institute for Astronomy, the Queen’s University Belfast, the Space Telescope Science Institute, the South African Astronomical Observatory, and The Millennium Institute of Astrophysics (MAS), Chile.
      
      The ZTF forced-photometry service was funded under the Heising-Simons Foundation grant $\#$12540303 (PI: Graham).

      This work makes use of observations from Neil Gehrels Swift Observatory (UVOT) and public data from the GALEX data archive.

\end{acknowledgements}

%
%

\bibliographystyle{aa} 
\bibliography{sn2020pvb_final} 

%

\begin{appendix} 
%
\section{Additional figures of \pvb.}

\begin{figure}[!ht]
\centering
\includegraphics[width=\columnwidth]{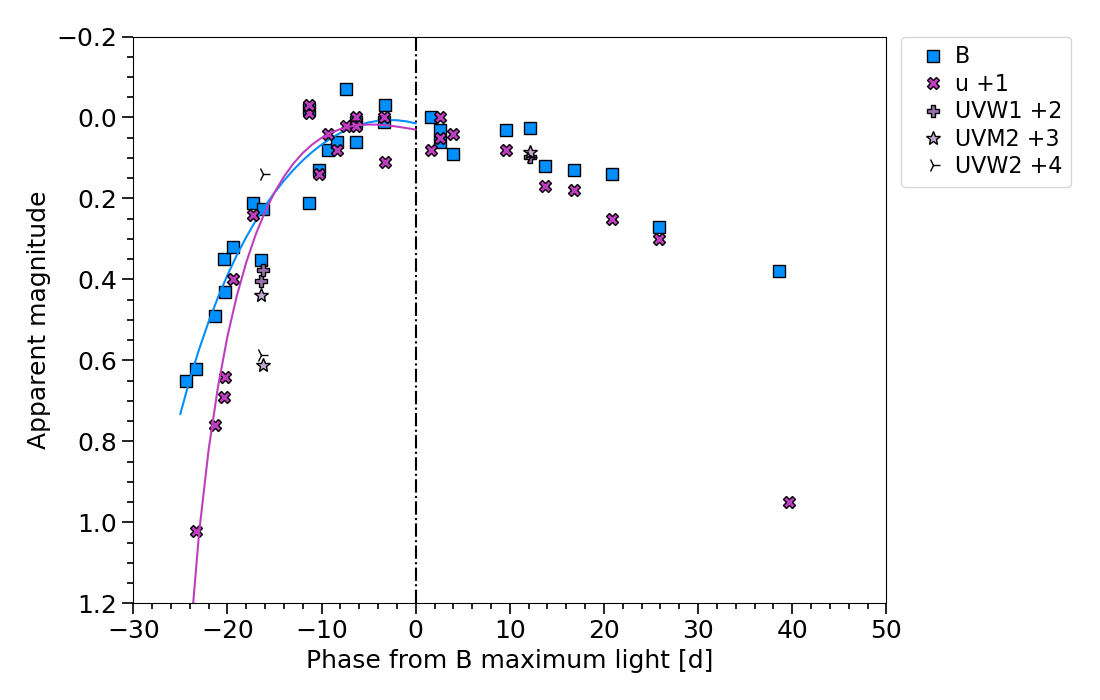} \hfill 
\caption{Bluest bands light curves of \pvb shifted at its maximum light. The solid line is the best-fit polynomial of the light curve rise. The dot-dashed vertical line indicates the $B$-band maximum light. }
\label{fig_bluebands}%
\end{figure}

\begin{figure}[!ht]
\centering
\includegraphics[width=\columnwidth]{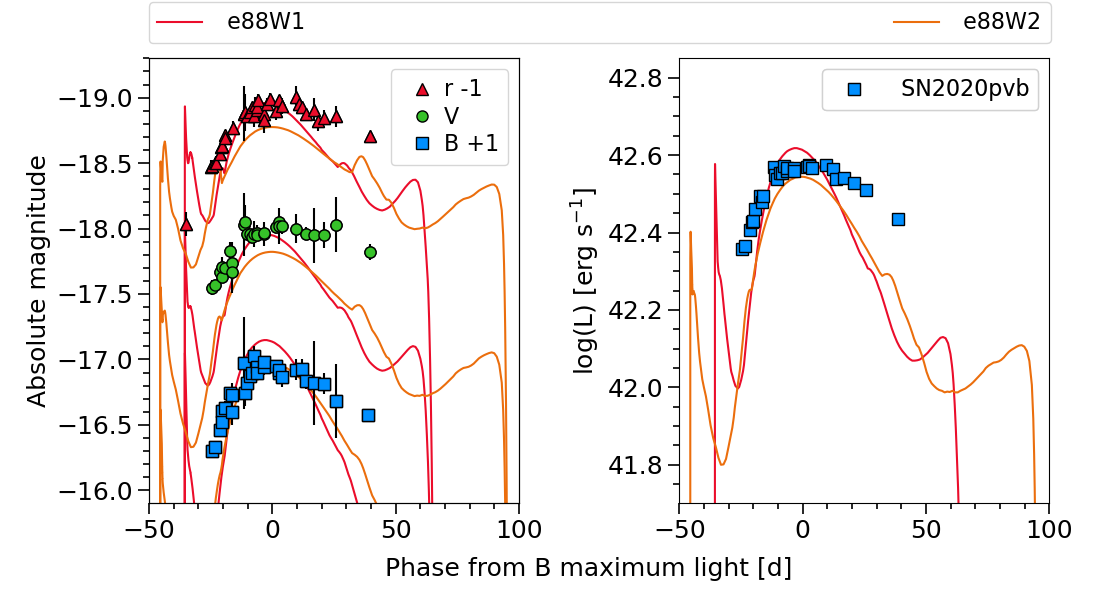} \hfill 
\caption{Absolute {\it BVr} {\it (left)} and {\it uUBVri} pseudo-bolometric {\it (right)} light curves of \pvb, compared with those for the ECSN model from \cite{kozyreva21} surrounded by a wind-like CSM.}
\label{fig_ECSNmodels}%
\end{figure}
\section{Tables of photometry and spectroscopy of \pvb.}

\begin{table*}
\caption{Basic information about the telescopes and instruments used (in alphabetical key order).}
\label{table_facilities}
\scalebox{0.90}{
\begin{tabular}{@{}lllll@{}}
\hline
Table Key & Telescope & Instrument & Pixel-scale & Location \\
& & & (arcsec pixel$^{-1}$) & \\
\hline
ALFOSC$^*$ & 2.56 m Nordic Optical Telescope & ALFOSC  & 0.19  &  RMO$^a$, Spain\\
ATLAS & 2x0.5 m Schmidt telescopes & CCD & 1.86 & Mount Haleakala Obs., USA \\
EFOSC2$^{**}$ & 3.58 m New Technology Telescope & EFOSC2 &  0.24 & ESO$^b$, La Silla Obs., Chile \\
EMIR & 10.40 m Gran Telescopio CANARIAS & EMIR  & 0.19  &  RMO, Spain\\
HST & 2.40 m Hubble Space Telescope & ACS/WFC & 0.05 & - \\
IO:O & 2.00 m Liverpool Telescope & IO:O   & 0.30  & RMO, Spain\\
LCO & 1.00 m LCO (CPT site) & Sinistro & 0.39 & LCO node at SAAO$^c$, South Africa\\
 & 1.00 m LCO (LSC site) & Sinistro & 0.39 & LCO node at CTIO$^d$, Chile\\
LRS & 3.58 m Telescopio Nazionale Galileo & LRS  & 0.25  &  RMO, Spain\\
NOTCam$^*$ & 2.56 m Nordic Optical Telescope & NOTCam & 0.23  & RMO, Spain\\
OSIRIS & 10.40 m Gran Telescopio CANARIAS & OSIRIS  & 0.25  &  RMO, Spain\\
PS1 & 1.80 m Pan-STARRS Telescope & GPC1 & 0.25  & Mount Haleakala Obs., USA \\
TRAPPIST & 0.60 m TRAPPIST-S Telescope & FLI ProLine & 0.65 & ESO, La Silla Obs., Chile\\
UVOT & 0.30 m Ritchey-Chretien UV/optical Telescope & Neil Gehrels Swift Obs. & 0.50 & - \\
WIFSIP & 1.20 m STELLA Telescope & WiFSIP & 0.32 & Iza\~{n}a Observatory, Spain \\
ZTF & 1.22 m Palomar Schmidt Telescope & 47-square-degree CCD & 1.00 & Palomar Obs., USA\\
\hline
\end{tabular}
}
\begin{flushleft}
$^*$ Data taken in the framework of the Nordic-optical-telescope Unbiased Transient Survey 2 (NUTS2) collaboration. \url{https://nuts.sn.ie/}\\
$^{**}$ Data taken in the framework of the extended Public ESO -European Southern Observatory- Spectroscopic Survey for Transient Objects (ePESSTO+). \url{https://www.pessto.org/}\\
$^a$ Roque de Los Muchachos Observatory.\\
$^b$ European Southern Observatory.\\
$^c$South African Astronomical Observatory.\\
$^d$Cerro Tololo Inter-American Observatory.\\
\end{flushleft}
\end{table*}


\begin{table*}
\caption{$UBV$ ({\sc Vega mag}) photometry of \pvb.}
\label{table_ph}
\begin{tabular}{@{}cccccccc@{}}
\hline 
Date & MJD & Phase$^a$  & U & B & V & Instrument key \\ 
 &  & (days) & (mag) & (mag) & (mag)  & \\ 
\hline 
2019-10-01 & 58757.70 &  -401.5 & $>  19.7 $&  - &  - & UVOT  \\ 
2019-10-04 & 58760.44 &  -398.7 & $>  21.2 $&  $>  21.1 $&  $>  20.7 $& UVOT  \\ 
2020-10-12 & 59134.80 &   -24.4 & - &  17.74 (0.02) &  17.39 (0.06) & LRS  \\ 
2020-10-13 & 59135.80 &   -23.4 & - &  17.71 (0.01) &  17.37 (0.03) & ALFOSC  \\ 
2020-10-15 & 59137.90 &   -21.3 & - &  17.58 (0.02) &  17.27 (0.05) & ALFOSC  \\ 
2020-10-16 & 59138.80 &   -20.4 & - &  17.44 (0.04) &  17.31 (0.03) & WIFSIP  \\ 
2020-10-16 & 59138.90 &   -20.3 & - &  17.52 (0.04) &  17.23 (0.07) & IO:O  \\ 
2020-10-17 & 59139.80 &   -19.4 & - &  17.41 (0.05) &  17.24 (0.04) & WIFSIP  \\ 
2020-10-19 & 59141.90 &   -17.3 & - &  17.30 (0.05) &  17.11 (0.07) & WIFSIP  \\ 
2020-10-20 & 59142.76 &   -16.4 & 16.50 (0.09) &  17.44 (0.10) &  17.20 (0.15) & UVOT  \\ 
2020-10-20 & 59142.96 &   -16.2 & 16.42 (0.09) &  17.32 (0.10) &  17.27 (0.16) & UVOT  \\ 
2020-10-25 & 59147.80 &   -11.4 & - &  17.07 (0.35) &  16.91 (0.24) & WIFSIP  \\ 
2020-10-25 & 59147.90 &   -11.3 & - &  17.30 (0.10) &  16.89 (0.13) & IO:O  \\ 
2020-10-26 & 59148.90 &   -10.3 & - &  17.22 (0.01) &  16.98 (0.02) & ALFOSC  \\ 
2020-10-27 & 59149.90 &    -9.3 & - &  17.17 (0.03) &  16.98 (0.03) & IO:O  \\ 
2020-10-28 & 59150.80 &    -8.4 & - &  17.15 (0.01) &  17.00 (0.04) & WIFSIP  \\ 
2020-10-29 & 59151.80 &    -7.4 & - &  17.02 (0.08) &  16.98 (0.10) & LCO  \\ 
2020-10-30 & 59152.80 &    -6.4 & - &  17.10 (0.01) &  16.99 (0.02) & WIFSIP  \\ 
2020-10-30 & 59152.80 &    -6.4 & - &  17.15 (0.04) &  16.97 (0.06) & IO:O  \\ 
2020-11-02 & 59155.80 &    -3.4 & - &  17.10 (0.06) &  16.98 (0.09) & LCO  \\ 
2020-11-02 & 59155.90 &    -3.3 & - &  17.06 (0.01) &  16.97 (0.03) & WIFSIP  \\ 
2020-11-07 & 59160.80 &     1.6 & - &  17.09 (0.01) &  16.93 (0.03) & WIFSIP  \\ 
2020-11-08 & 59161.80 &     2.6 & - &  17.12 (0.03) &  16.89 (0.05) & ALFOSC  \\ 
2020-11-08 & 59161.80 &     2.6 & - &  17.15 (0.08) &  16.92 (0.14) & LCO  \\ 
2020-11-10 & 59163.10 &     3.9 & - &  17.18 (0.07) &  16.92 (0.06) & LCO  \\ 
2020-11-15 & 59168.80 &     9.6 & - &  17.12 (0.08) &  16.94 (0.11) & LCO  \\ 
2020-11-18 & 59171.35 &    12.2 & 16.30 (0.07) &  17.11 (0.07) &  - & UVOT  \\ 
2020-11-19 & 59172.90 &    13.7 & - &  17.21 (0.06) &  16.98 (0.05) & ALFOSC  \\ 
2020-11-23 & 59176.00 &    16.8 & - &  17.22 (0.32) &  16.99 (0.21) & LCO  \\ 
2020-11-27 & 59180.00 &    20.8 & - &  17.23 (0.08) &  16.99 (0.10) & LCO  \\ 
2020-12-02 & 59185.00 &    25.8 & - &  17.36 (0.28) &  16.91 (0.21) & LCO  \\ 
2020-12-14 & 59197.80 &    38.6 & - &  17.47 (0.05) &  - & ALFOSC  \\ 
2020-12-15 & 59198.80 &    39.6 & - &  - &  17.12 (0.06) & ALFOSC  \\ 
2021-04-12 & 59316.40 &   157.2 & - &  $>  22.5 $&  $>  21.7 $& EFOSC2  \\ 
2021-05-14 & 59348.30 &   189.1 & - &  - &  $>  21.1 $& TRAPPIST  \\ 
2021-05-22 & 59356.40 &   197.2 & - &  - &  $>  21.5 $& TRAPPIST  \\ 
2021-06-02 & 59367.40 &   208.2 & - &  - &  $>  20.4 $& TRAPPIST  \\ 
2021-07-06 & 59401.09 &   241.9 & - &  $>  22.4 $&  $>  21.6 $& LRS  \\ 
\hline  
\end{tabular}
\begin{flushleft}
$^a$ Phases are relative to $B$ maximum light, MJD = 59159.18 $\pm$ 0.50.\\ 
\end{flushleft}
\end{table*}

\begin{table*}
\caption{$ugriz$ ({\sc AB mag}) photometry of \pvb.}
\label{table_SLph}
\begin{tabular}{@{}ccccccccc@{}}
\hline  
Date & MJD & Phase$^a$ & u & g & r & i  & z & Instrument key\\ 
 &  & (days) & (mag) & (mag) & (mag) & (mag)  & (mag) &  \\ 
\hline  
2018-11-02 & 58424.11 & -735.1 & - & - &  $>$ 18.9 & - & - & ZTF \\
2018-11-03 & 58425.11 & -734.1 & - & - &  $>$ 20.1 & - & - & ZTF \\
2018-11-09 & 58431.10 & -728.1 & - & $>$ 19.7 & $>$ 19.8 & - & - & ZTF \\ 
2019-06-11 & 58645.44 & -513.7 & - & $>$ 20.5 & - & - & - & ZTF \\
2019-06-23 & 58657.42 & -501.8 & - & - &  $>$ 19.8 & - & - & ZTF \\
2019-06-27 & 58661.48 & -497.7 & - & $>$ 19.8 & $>$ 20.4 & - & - & ZTF \\ 
2019-07-03 & 58667.44 & -491.7 & - & $>$ 20.7 & $>$ 19.7 & - & - & ZTF \\
2019-07-06 & 58670.44 & -488.7 & - & $>$ 21.0 & - & - & - & ZTF \\
2019-07-09 & 58673.48 & -485.7 & - & $>$ 20.2 & $>$ 20.5 & - & - & ZTF \\
2019-07-12 & 58676.48 & -482.7 & - & - &  $>$ 20.3 & - & - & ZTF \\
2019-07-16 & 58680.44 & -478.7 & - & $>$ 19.1 & - & - & - & ZTF \\
2019-07-20 & 58684.44 & -474.7 & - & - &  $>$ 19.3 & - & - & ZTF \\
2019-07-26 & 58690.44 & -468.7 & - & $>$ 20.2 & $>$ 20.5 & - & - & ZTF \\ 
2019-07-27 & 58691.44 & -467.7 & - & $>$ 20.0 & $>$ 20.5 & - & - & ZTF \\ 
2019-07-28 & 58692.36 & -466.8 & - & $>$ 19.1 & $>$ 19.0 & - & - & ZTF \\ 
2019-07-29 & 58693.33 & -465.9 & - & $>$ 20.6 & $>$ 20.4 & - & - & ZTF \\ 
2019-07-30 & 58694.36 & -464.8 & - & - & $>$ 20.6 & - & - & ZTF \\ 
2019-07-31 & 58695.39 & -463.8 & - & $>$ 21.1 &  $>$ 21.0 & - & - & ZTF \\
2019-08-01 & 58696.37 & -462.8 & - & - &  $>$ 20.5 & - & - & ZTF \\
2019-08-03 & 58698.33 & -460.9 & - & $>$ 17.9 & $>$ 19.8 & - & - & ZTF \\ 
2019-08-06 & 58701.33 & -457.9 & - & $>$ 20.9 & $>$ 20.2 & - & - & ZTF \\ 
2019-08-11 & 58706.39 & -452.8 & - & - &  $>$ 19.6 & - & - & ZTF \\
2019-08-20 & 58715.31 & -443.9 & - & - &  $>$ 19.8 & - & - & ZTF \\
2019-08-23 & 58718.32 & -440.9 & - & - &  $>$ 20.0 & - & - & ZTF \\
2019-08-27 & 58722.30 & -436.9 & - & - &  $>$ 20.2 & - & - & ZTF \\
2019-08-30 & 58725.32 & -433.9 & - & $>$ 20.4 & - & - & - & ZTF \\
2019-09-07 & 58733.30 & -425.9 & - & $>$ 20.1 & - & - & - & ZTF \\
2019-09-17 & 58743.21 & -416.0 & - & $>$ 19.7 &  $>$ 19.5 & - & - & ZTF \\
2019-09-21 & 58747.30 & -411.9 & - & $>$ 19.8 &  $>$ 19.7 & - & - & ZTF \\
2019-09-24 & 58750.25 & -408.9 & - & - & $>$ 19.7 & - & - & ZTF \\
2019-10-01 & 58757.16 & -402.0 & - & $>$ 20.6 & $>$ 20.3 & - & - & ZTF \\
2019-10-04 & 58760.25 & -398.9 & - & - & $>$ 19.9 & - & - & ZTF \\
2019-10-10 & 58766.19 & -393.0 & - & $>$ 19.1 & $>$ 19.1 & - & - & ZTF \\
2019-10-13 & 58769.19 & -390.0 & - & $>$ 19.3 & $>$ 19.3 & - & - & ZTF \\
2019-10-17 & 58773.18 & -386.0 & - & $>$ 19.1 & $>$ 18.4 & - & - & ZTF \\
2019-10-20 & 58776.18 & -383.0 & - & $>$ 20.1 & $>$ 19.6 & - & - & ZTF \\
2019-10-26 & 58782.18 & -377.0 & - & - & $>$ 19.7 & - & - & ZTF \\
2019-10-29 & 58785.18 & -374.0 & - & $>$ 20.0 & $>$ 19.8 & - & - & ZTF \\
2019-11-02 & 58789.18 & -370.0 & - & $>$ 19.8 & - & - & - & ZTF \\
2019-11-07 & 58794.13 & -365.1 & - & $>$ 19.4 & $>$ 19.8 & - & - & ZTF \\
2019-11-10 & 58797.13 & -362.1 & - & $>$ 19.4 & $>$ 19.7 & - & - & ZTF \\
2020-07-10 & 59040.46 & -118.7 & - & $>$ 19.9 & $>$ 20.4 & - & - & ZTF \\
2020-07-13 & 59043.36 & -115.8 & - & $>$ 20.5 & - & - & - & ZTF \\
2020-10-02 & 59124.23 & -35.0 & - & - & 17.85 (0.05)  & - & - & ZTF \\
2020-10-07 & 59129.17 & -30.0 & - & $>$ 16.7 & - & - & - & ZTF \\
2020-10-12 & 59134.17 & -25.0 & - & - & 17.45 (0.02)  & - & - & ZTF \\
2020-10-12 & 59134.80 &   -24.4 & - &  17.49 (0.03) &  17.40 (0.04) &  17.52 (0.04) &  17.62 (0.07) & LRS  \\ 
2020-10-13 & 59135.80 &   -23.4 & 17.94 (0.04) &  17.44 (0.03) &  17.38 (0.02) &  17.48 (0.02) &  17.55 (0.05) & ALFOSC  \\ 
2020-10-14 & 59136.13 & -23.1 & - & 17.51 (0.02) & 17.39 (0.02) & - & - & ZTF \\	 
2020-10-15 & 59137.90 &   -21.3 & 17.68 (0.12) &  17.34 (0.02) &  17.31 (0.03) &  17.42 (0.03) &  17.51 (0.07) & ALFOSC  \\ 
2020-10-16 & 59138.19 & -21.0 & - & 17.42 (0.02) & 17.21 (0.02) & - & - & ZTF \\
2020-10-16 & 59138.80 &   -20.4 & 17.61 (0.07) &  17.29 (0.03) &  17.25 (0.03) &  17.40 (0.05) &  17.41 (0.08) & WIFSIP  \\ 
2020-10-16 & 59138.90 &   -20.3 & 17.56 (0.06) &  17.37 (0.03) &  17.25 (0.03) &  17.35 (0.03) &  17.45 (0.07) & IO:O  \\ 
2020-10-17 & 59139.80 &   -19.4 & 17.32 (0.10) &  17.24 (0.03) &  17.16 (0.03) &  17.33 (0.05) &  17.41 (0.09) & WIFSIP  \\ 
2020-10-18 & 59140.15 & -19.0 & - & 17.29 (0.01) & 17.17 (0.01) & - & - & ZTF \\
2020-10-19 & 59141.90 &   -17.3 & 17.16 (0.19) &  - &  - &  - &  - & WIFSIP  \\ 
2020-10-21 & 59143.17 & -16.0 & - & 17.12 (0.01) & 17.06 (0.01) & - & - & ZTF \\
2020-10-25 & 59147.80 &   -11.4 & 16.91 (0.27) &  17.08 (0.37) &  17.00 (0.21) &  17.22 (0.18) &  17.19 (0.27) & WIFSIP  \\ 
\hline  
\end{tabular}
\begin{flushleft}
$^a$ Phases are relative to $B$ maximum light, MJD = 59159.18 $\pm$ 0.50.\\ 
\end{flushleft}
\end{table*}

\begin{table*}
\contcaption{}
\begin{tabular}{@{}ccccccccc@{}}
\hline  
Date & MJD & Phase$^a$ & u & g & r & i  & z & Instrument key\\ 
 &  & (days) & (mag) & (mag) & (mag) & (mag)  & (mag) &  \\ 
\hline  
2020-10-25 & 59147.90 &   -11.3 & 16.89 (0.15) &  17.05 (0.17) &  16.99 (0.14) &  17.18 (0.17) &  17.26 (0.13) & IO:O  \\ 
2020-10-26 & 59148.90 &   -10.3 & 17.06 (0.09) &  17.03 (0.04) &  17.02 (0.03) &  17.14 (0.01) &  17.26 (0.03) & ALFOSC  \\ 
2020-10-27 & 59149.90 &    -9.3 & 16.96 (0.06) &  17.03 (0.03) &  16.99 (0.03) &  17.10 (0.04) &  17.20 (0.05) & IO:O  \\ 
2020-10-28 & 59150.15 & -9.0  & - & 17.06 (0.02) & 17.00 (0.03) & - & - & ZTF \\
2020-10-28 & 59150.80 &    -8.4 & 17.00 (0.08) &  16.99 (0.02) &  16.95 (0.01) &  17.10 (0.02) &  17.21 (0.04) & WIFSIP  \\ 
2020-10-29 & 59151.16 & -8.0  & - & 17.07 (0.03)  & - & - & - & ZTF \\
2020-10-29 & 59151.80 &    -7.4 & 16.94 (0.05) &  17.00 (0.07) &  17.02 (0.08) &  17.08 (0.08) &  17.19 (0.14) & LCO  \\ 
2020-10-30 & 59152.11 & -7.1  & - & 17.02 (0.02) & 16.93 (0.02) & - & - & ZTF \\
2020-10-30 & 59152.80 &    -6.4 & 16.94 (0.05) &  17.01 (0.04) &  16.97 (0.02) &  17.09 (0.03) &  17.18 (0.03) & WIFSIP  \\ 
2020-10-30 & 59152.80 &    -6.4 & 16.92 (0.03) &  17.00 (0.02) &  16.95 (0.02) &  17.11 (0.03) &  17.19 (0.05) & IO:O  \\ 
2020-10-31 & 59153.13 & -6.1  & - & 17.00 (0.02) & 16.91 (0.02) & - & - & ZTF \\
2020-11-02 & 59155.80 &    -3.4 & 16.92 (0.07) &  16.99 (0.06) &  17.00 (0.06) &  17.07 (0.08) &  17.02 (0.12) & LCO  \\ 
2020-11-02 & 59155.90 &    -3.3 & 17.03 (0.07) &  16.97 (0.10) &  17.05 (0.10) &  17.13 (0.20) &  17.07 (0.18) & WIFSIP  \\ 
2020-11-04 & 59157.17 & -2.0  & - & 17.00 (0.02) & 16.92 (0.01) & - & - & ZTF \\
2020-11-05 & 59158.18 & -1.0  & - & 17.04 (0.02) & 16.91 (0.01) & - & - & ZTF \\
2020-11-07 & 59160.80 &     1.6 & 17.00 (0.01) &  16.93 (0.01) &  - &  - &  - & WIFSIP  \\ 
2020-11-07 & 59160.80 &     1.6 & - &  - &  16.98 (0.07) &  - &  - & OSIRIS  \\ 
2020-11-08 & 59161.80 &     2.6 & 16.92 (0.14) &  16.93 (0.03) &  16.90 (0.04) &  17.02 (0.05) &  17.11 (0.06) & ALFOSC  \\ 
2020-11-08 & 59161.80 &     2.6 & 16.97 (0.06) &  - &  - &  - &  - & LCO  \\ 
2020-11-10 & 59163.10 &     3.9 & 16.96 (0.05) &  16.95 (0.04) &  16.94 (0.06) &  17.05 (0.07) &  17.06 (0.09) & LCO  \\ 
2020-11-12 & 59165.15 & 6.0   & - & 16.98	(0.01)  & - & - & - & ZTF \\
2020-11-15 & 59168.80 &     9.6 & 17.00 (0.05) &  16.94 (0.07) &  16.87 (0.08) &  17.06 (0.10) &  17.09 (0.17) & LCO  \\ 
2020-11-16 & 59169.11 & 9.9   & - & 17.08	(0.02)  & - & - & - & ZTF \\
2020-11-17 & 59170.13 & 10.9 & - & - & 16.92 (0.01) & - & - & ZTF \\	
2020-11-18 & 59171.07 & 11.9 & - & - & 16.91 (0.02) & - & - & ZTF \\					
2020-11-19 & 59172.90 &    13.7 & 17.09 (0.06) &  17.02 (0.03) &  17.00 (0.04) &  17.14 (0.04) &  17.23 (0.09) & ALFOSC  \\ 
2020-11-23 & 59176.00 &    16.8 & 17.10 (0.10) &  17.02 (0.08) &  16.97 (0.09) &  17.11 (0.13) &  17.24 (0.24) & LCO  \\ 
2020-11-24 & 59177.80 &    18.6 & - &  - &  17.06 (0.07) &  - &  - & OSIRIS  \\ 
2020-11-27 & 59180.00 &    20.8 & 17.17 (0.04) &  17.10 (0.06) &  17.03 (0.06) &  17.16 (0.09) &  17.15 (0.14) & LCO  \\ 
2020-12-02 & 59185.00 &    25.8 & 17.22 (0.09) &  17.14 (0.09) &  17.02 (0.08) &  17.15 (0.14) &  17.16 (0.20) & LCO  \\ 
2020-12-15 & 59198.80 &    39.6 & 17.87 (0.04) &  17.30 (0.03) &  17.17 (0.02) &  17.27 (0.03) &  17.35 (0.04) & ALFOSC  \\ 
2021-04-12 & 59316.40 &   157.2 & - &  - &  $>  22.9 $&  $>  22.5 $&  - & EFOSC2$^b$  \\ 
2021-05-02 & 59336.40 &   177.2 & - &  - &  - &  $>  21.6 $&  - & TRAPPIST$^b$  \\ 
2021-05-14 & 59348.30 &   189.1 & - &  - &  $>  21.6 $&  - &  - & TRAPPIST$^b$  \\ 
2021-06-02 & 59367.40 &   208.2 & - &  - &  $>  21.0 $&  - &  - & TRAPPIST$^b$  \\ 
2021-06-03 & 59368.20 &   209.0 & - &  - &  $>  20.5 $&  - &  - & OSIRIS  \\ 
2021-06-20 & 59385.42 & 226.2 & - & $>$ 20.5 & - & - & - & ZTF \\
2021-06-22 & 59387.20 &   228.0 & - &  - &  $>  21.5 $&  - &  - & TRAPPIST$^b$  \\ 
2021-06-22 & 59387.42 & 228.2 & - & - & $>$ 18.4 & - & - & ZTF \\
2021-06-29 & 59394.42 & 235.2 & - & - & $>$ 19.6 & - & - & ZTF \\
2021-07-01 & 59396.40 & 237.2 & - & - & $>$ 20.1 & - & - & ZTF \\
2021-07-04 & 59399.38 & 240.2 & - & - & $>$ 20.4 & - & - & ZTF \\
2021-07-06 & 59401.10 &   241.9 & - &  $>  21.1 $&  $>  20.9 $&  $>  21.0 $&  $>  20.9 $& LRS  \\ 
2021-07-06 & 59401.39 & 242.2 & - & $>$ 20.4 & - & - & - & ZTF \\
2021-07-08 & 59403.39 & 244.2 & - & $>$ 20.5 & - & - & - & ZTF \\
2021-07-10 & 59405.39 & 246.2 & - & - & $>$ 20.2 & - & - & ZTF \\
2021-07-14 & 59409.40 & 250.2 & - & $>$ 20.2 & - & - & - & ZTF \\
2021-07-17 & 59412.40 & 253.2 & - & - & $>$ 20.3 & - & - & ZTF \\
2021-07-20 & 59415.42 & 256.2 & - & $>$ 20.5 & $>$ 19.9 & - & - & ZTF \\
2021-07-27 & 59422.44 & 263.3 & - & $>$ 19.3 & - & - & - & ZTF \\	
2021-07-29 & 59424.42 & 265.2 & - & $>$ 19.9 & - & - & - & ZTF \\	
2021-07-31 & 59426.35 & 267.2 & - & $>$ 20.4 & $>$ 20.4 & - & - & ZTF \\
2021-08-02 & 59428.38 & 269.2 & - & - & $>$ 20.1 & - & - & ZTF \\
2021-08-04 & 59430.34 & 271.2 & - & - & $>$ 20.6 & - & - & ZTF \\
2021-08-08 & 59434.28 & 275.1 & - & - & $>$ 20.7 & - & - & ZTF \\
2021-08-09 & 59435.38 & 276.2 & - & $>$ 20.4 & - & - & - & ZTF \\
2021-08-12 & 59438.32 & 279.1 & - & - & $>$ 20.0 & - & - & ZTF \\
2021-08-14 & 59440.27 & 281.1 & - & $>$ 20.5 & $>$ 20.3 & - & - & ZTF \\ 
\hline  
\end{tabular}
\begin{flushleft}
$^a$ Phases are relative to $B$ maximum light, MJD = 59159.18 $\pm$ 0.50.\\ 
$^b$ Vega mag converted to AB mag using \cite{blanton07}. 
\end{flushleft}
\end{table*}

\begin{table*}
\contcaption{}
\begin{tabular}{@{}ccccccccc@{}}
\hline  
Date & MJD & Phase$^a$ & u & g & r & i  & z & Instrument key\\ 
 &  & (days) & (mag) & (mag) & (mag) & (mag)  & (mag) &  \\ 
\hline  
2021-08-17 & 59443.31 & 284.1 & - & $>$ 19.8 & $>$ 19.6 & - & - & ZTF \\ 
2021-08-23 & 59449.25 & 290.1 & - & $>$ 19.1 & - & - & - & ZTF \\	
2021-08-25 & 59451.23 & 292.0 & - & $>$ 19.4 & $>$ 19.4 & - & - & ZTF \\
2021-08-27 & 59453.36 & 294.2 & - & $>$ 19.9 & $>$ 20.1 & - & - & ZTF \\
2021-08-29 & 59455.23 & 296.0 & - & $>$ 20.8 & $>$ 20.3 & - & - & ZTF \\
2021-09-02 & 59459.23 & 300.0 & - & - & $>$ 20.5 & - & - & ZTF \\		
2021-09-04 & 59461.21 & 302.0 & - & $>$ 20.4 & - & - & - & ZTF \\
2021-09-06 & 59463.23 & 304.0 & - & $>$ 20.6 & $>$ 20.3 & - & - & ZTF \\
2021-09-08 & 59465.21 & 306.0 & - & $>$ 20.5 & $>$ 20.6 & - & - & ZTF \\
2021-09-10 & 59467.19 & 308.0 & - & $>$ 20.3 & $>$ 20.5 & - & - & ZTF \\
2021-09-12 & 59469.23 & 310.0 & - & $>$ 20.6 & $>$ 20.5 & - & - & ZTF \\
2021-09-14 & 59471.25 & 312.1 & - & $>$ 19.7 & $>$ 19.7 & - & - & ZTF \\
2021-09-19 & 59476.23 & 317.0 & - & $>$ 19.2 & $>$ 19.3 & - & - & ZTF \\
2021-09-21 & 59478.23 & 319.0 & - & - & $>$ 19.4 & - & - & ZTF \\
2021-09-23 & 59480.21 & 321.0 & - & - & $>$ 20.0 & - & - & ZTF \\
2021-09-27 & 59484.25 & 325.1 & - & - & $>$ 20.3 & - & - & ZTF \\
2021-09-30 & 59487.23 & 328.0 & - & - & $>$ 20.6 & - & - & ZTF \\
2021-10-02 & 59489.25 & 330.1 & - & - & $>$ 20.0 & - & - & ZTF \\
2021-10-04 & 59491.25 & 332.1 & - & $>$ 19.0 & $>$ 20.2 & - & - & ZTF \\
2021-10-10 & 59497.21 & 338.0 & - & - & $>$ 19.4 & - & - & ZTF \\
2021-10-17 & 59504.19 & 345.0 & - & $>$ 20.1 & - & - & - & ZTF \\ 			
2021-10-23 & 59510.19 & 351.0 & - & $>$ 18.0 & - & - & - & ZTF \\	
2021-10-25 & 59512.15 & 353.0 & - & - & $>$ 20.5 & - & - & ZTF \\
2021-11-03 & 59521.11 & 361.9 & - & $>$ 20.5 & - & - & - & ZTF \\			
2021-11-06 & 59524.11 & 364.9 & - & $>$ 20.8 & $>$ 20.1 & - & - & ZTF \\
2021-11-07 & 59525.11 & 365.9 & - & $>$ 20.7 & - & - & - & ZTF \\			
2021-11-19 & 59537.11 & 377.9 & - & $>$ 16.9 & - & - & - & ZTF \\			
2022-06-19 & 59749.42 & 590.2 & - & $>$ 19.8 & - & - & - & ZTF \\			
2022-06-21 & 59751.42 & 592.2 & - & - & $>$ 20.4 & - & - & ZTF \\
2022-06-23 & 59753.42 & 594.2 & - & $>$ 20.6 & - & - & - & ZTF \\		
2022-07-03 & 59763.40 & 604.2 & - & $>$ 20.6 & $>$ 20.6 & - & - & ZTF \\
2022-07-05 & 59765.40 & 606.2 & - & $>$ 20.9 & $>$ 20.8 & - & - & ZTF \\
2022-07-08 & 59768.38 & 609.2 & - & $>$ 20.9 & - & - & - & ZTF \\			
2022-07-10 & 59770.40 & 611.2 & - & $>$ 20.9 & - & - & - & ZTF \\			
2022-07-12 & 59772.36 & 613.2 & - & - & $>$ 19.7 & - & - & ZTF \\
2022-07-18 & 59778.38 & 619.2 & - & - & $>$ 19.1 & - & - & ZTF \\
2022-07-20 & 59780.44 & 621.3 & - & $>$ 20.0 & - & - & - & ZTF \\			
2022-07-22 & 59782.44 & 623.3 & - & $>$ 19.7 & - & - & - & ZTF \\			
2022-07-24 & 59784.36 & 625.2 & - & $>$ 20.9 & - & - & - & ZTF \\			
2022-07-26 & 59786.33 & 627.2 & - & $>$ 20.5 & $>$ 21.0 & - & - & ZTF \\
2022-07-28 & 59788.38 & 629.2 & - & - & $>$ 20.5 & - & - & ZTF \\
2022-07-30 & 59790.33 & 631.2 & - & $>$ 19.6 & $>$ 19.2 & - & - & ZTF \\
2022-08-03 & 59794.42 & 635.2 & - & $>$ 20.0 & $>$ 20.4 & - & - & ZTF \\
2022-08-06 & 59797.38 & 638.2 & - & $>$ 20.6 & $>$ 20.7 & - & - & ZTF \\
2022-08-14 & 59805.32 & 646.1 & - & - & $>$ 18.7 & - & - & ZTF \\
2022-08-16 & 59807.31 & 648.1 & - & - & $>$ 20.2 & - & - & ZTF \\
2022-08-18 & 59809.33 & 650.2 & - & $>$ 20.0 & $>$ 20.1 & - & - & ZTF \\
2022-08-20 & 59811.34 & 652.2 & - & - & $>$ 20.4 & - & - & ZTF \\
2022-08-23 & 59814.27 & 655.1 & - & $>$ 20.6 & $>$ 20.6 & - & - & ZTF \\
2022-08-25 & 59816.26 & 657.1 & - & $>$ 20.7 & $>$ 20.4 & - & - & ZTF \\
2022-08-31 & 59822.32 & 663.1 & - & $>$ 20.3 & $>$ 20.2 & - & - & ZTF \\
2022-09-02 & 59824.21 & 665.0 & - & $>$ 20.6 & $>$ 20.3 & - & - & ZTF \\
2022-09-04 & 59826.27 & 667.1 & - & - & $>$ 20.2 & - & - & ZTF \\
2022-09-06 & 59828.23 & 669.0 & - & $>$ 19.4 & - & - & - & ZTF \\			
2022-09-17 & 59839.23 & 680.0 & - & $>$ 20.8 & - & - & - & ZTF \\			
2022-09-19 & 59841.23 & 682.0 & - & $>$ 20.9 & - & - & - & ZTF \\			
2022-09-21 & 59843.19 & 684.0 & - & $>$ 20.6 & - & - & - & ZTF \\			
2022-09-23 & 59845.21 & 686.0 & - & $>$ 20.8 & $>$ 20.4 & - & - & ZTF \\
2022-09-25 & 59847.27 & 688.1 & - & $>$ 20.2 & - & - & - & ZTF \\			
\hline  
\end{tabular}
\begin{flushleft}
$^a$ Phases are relative to $B$ maximum light, MJD = 59159.18 $\pm$ 0.50.\\ 
\end{flushleft}
\end{table*}

\begin{table*}
\contcaption{}
\begin{tabular}{@{}ccccccccc@{}}
\hline  
Date & MJD & Phase$^a$ & u & g & r & i  & z & Instrument key\\ 
 &  & (days) & (mag) & (mag) & (mag) & (mag)  & (mag) &  \\ 
\hline  
2022-09-27 & 59849.21 & 690.0 & - & $>$ 20.5 & $>$ 20.3 & - & - & ZTF \\
2022-09-29 & 59851.19 & 692.0 & - & $>$ 20.4 & $>$ 20.3 & - & - & ZTF \\
2022-09-30 & 59852.15 & 693.0 & - & $>$ 20.5 & $>$ 20.5 & - & - & ZTF \\
2022-10-02 & 59854.22 & 695.0 & - & $>$ 20.7 & - & - & - & ZTF \\			
2022-10-07 & 59859.19 & 700.0 & - & $>$ 19.5 & $>$ 19.6 & - & - & ZTF \\
2022-10-09 & 59861.19 & 702.0 & - & $>$ 19.3 & - & - & - & ZTF \\	
2022-10-11 & 59863.21 & 704.0 & - & - & $>$ 20.0 & - & - & ZTF \\
2022-10-13 & 59865.19 & 706.0 & - & $>$ 20.0 & $>$ 20.2 & - & - & ZTF \\
2022-10-15 & 59867.21 & 708.0 & - & $>$ 20.2 & $>$ 20.7 & - & - & ZTF \\
2022-10-18 & 59870.13 & 710.9 & - & $>$ 20.5 & - & - & - & ZTF \\			
2022-10-20 & 59872.17 & 713.0 & - & $>$ 20.2 & - & - & - & ZTF \\			
2022-10-22 & 59874.17 & 715.0 & - & $>$ 20.7 & $>$ 20.5 & - & - & ZTF \\
2022-10-25 & 59877.15 & 718.0 & - & - & $>$ 20.3 & - & - & ZTF \\
2022-10-27 & 59879.20 & 720.0 & - & $>$ 20.0 & $>$ 20.3 & - & - & ZTF \\
2022-10-29 & 59881.11 & 721.9 & - & $>$ 20.3 & - & - & - & ZTF \\			
2022-10-31 & 59883.15 & 724.0 & - & $>$ 20.0 & $>$ 20.0 & - & - & ZTF \\
2022-11-04 & 59887.13 & 728.0 & - & - & $>$ 19.5 & - & - & ZTF \\
2022-11-05 & 59888.11 & 728.9 & - & - & $>$ 18.9 & - & - & ZTF \\
2022-11-07 & 59890.13 & 730.9 & - & $>$ 19.7 & - & - & - & ZTF \\			
2022-11-12 & 59895.13 & 735.9 & - & - & $>$ 20.1 & - & - & ZTF \\
2022-11-14 & 59897.13 & 737.9 & - & $>$ 20.0 & - & - & - & ZTF \\			
2022-11-15 & 59898.10 & 738.9 & - & $>$ 20.4 & - & - & - & ZTF \\			
2022-11-16 & 59899.11 & 739.9 & - & - & $>$ 19.8 & - & - & ZTF \\
2022-11-17 & 59900.13 & 740.9 & - & $>$ 19.2 & $>$ 19.4 & - & - & ZTF \\
2022-11-18 & 59901.13 & 741.9 & - & - & $>$ 16.6 & - & - & ZTF \\
2022-11-19 & 59902.13 & 742.9 & - & - & $>$ 19.6 & - & - & ZTF \\
\hline  
\end{tabular}
\begin{flushleft}
$^a$ Phases are relative to $B$ maximum light, MJD = 59159.18 $\pm$ 0.50.\\ 
\end{flushleft}
\end{table*}

\begin{table*}
\caption{$JHK$ ({\sc Vega mag}) photometry of \pvb.}
\label{table_NIRph}
\begin{tabular}{@{}ccccccc@{}}
\hline  
Date & MJD & Phase$^a$ & J & H & K & Instrument key\\ 
 &  & (days) & (mag) & (mag)  & (mag) &  \\ 
\hline  
2020-10-26 & 59148.86 & -10.3 & 16.65 (0.07) & -  & - & EMIR \\  
2020-11-20 & 59173.81 & 14.6 & 16.55 (0.02) & 16.12 (0.03) & 16.12 (0.03) & NOTCam \\  
2020-12-16 & 59199.80 & 40.6 & 16.54 (0.02) &  - &  - & NOTCam  \\ 
\hline  
\end{tabular}
\begin{flushleft}
$^a$ Phases are relative to $B$ maximum light, MJD = 59159.18 $\pm$ 0.50.\\ 
\end{flushleft}
\end{table*}

\begin{table*}
\caption{PS1 $w$-band ({\sc AB mag}) photometry of \pvb.}
\label{table_PSph}
\begin{tabular}{@{}ccccccc@{}}
\hline  
Date & MJD & Phase$^a$ & PS1.w & Instrument key\\ 
 &  & (days) & (mag)  &  \\ 
\hline  
2020-05-25 & 58994.56 &  -164.6 & $>  22.3 $& PS1  \\ 
2020-05-31 & 59000.57 &  -158.6 & $>  22.7 $& PS1  \\ 
2020-06-15 & 59015.54 &  -143.6 & $>  23.1 $& PS1  \\ 
2020-06-20 & 59020.54 &  -138.6 & $>  22.6 $& PS1  \\ 
2020-06-24 & 59024.50 &  -134.7 & $>  22.5 $& PS1  \\ 
2020-07-18 & 59048.42 &  -110.8 & 21.04 (0.18) & PS1  \\ 
2020-08-23 & 59084.34 &   -74.8 & $>  22.6 $& PS1  \\ 
2020-09-10 & 59102.31 &   -56.9 & 19.85 (0.09) & PS1  \\ 
2020-09-15 & 59107.25 &   -51.9 & 19.14 (0.05) & PS1  \\ 
2021-06-06 & 59371.54 &   212.4 & $>  22.2 $& PS1  \\ 
2021-07-07 & 59402.48 &   243.3 & $>  23.2 $& PS1 \\
2021-10-01 & 59488.25 &   329.1	& $>  22.2 $& PS1 \\ 
\hline  
\end{tabular}
\begin{flushleft}
$^a$ Phases are relative to $B$ maximum light, MJD = 59159.18 $\pm$ 0.50.\\ 
\end{flushleft}
\end{table*}

\begin{table*}
\caption{$cyan,orange$ ({\sc AB mag}) photometry of \pvb.}
\label{table_ATLASph}
\begin{tabular}{@{}ccccccc@{}}
\hline  
Date & MJD & Phase$^a$ & cyan & orange & Instrument key\\ 
 &  & (days) & (mag)  & (mag) &  \\ 
\hline  
2015-08-10 & 57244.42 & -1914.8 & $>  19.5 $&  - & ATLAS  \\ 
2015-10-01 & 57296.64 & -1862.5 & - &  $>  18.8 $& ATLAS  \\ 
2015-10-13 & 57308.32 & -1850.9 & $>  19.1 $&  - & ATLAS  \\ 
2015-10-30 & 57325.24 & -1833.9 & - &  $>  18.4 $& ATLAS  \\ 
2015-11-07 & 57333.22 & -1826.0 & $>  19.4 $&  - & ATLAS  \\ 
2016-05-03 & 57511.60 & -1647.6 & $>  19.4 $&  - & ATLAS  \\ 
2016-05-12 & 57520.59 & -1638.6 & $>  19.5 $&  - & ATLAS  \\ 
2016-06-28 & 57567.51 & -1591.7 & - &  $>  19.0 $& ATLAS  \\ 
2016-08-02 & 57602.43 & -1556.8 & $>  19.5 $&  - & ATLAS  \\ 
2016-09-08 & 57639.38 & -1519.8 & $>  19.7 $&  - & ATLAS  \\ 
2016-09-16 & 57647.36 & -1511.8 & - &  $>  18.2 $& ATLAS  \\ 
2016-10-02 & 57663.32 & -1495.9 & $>  19.4 $&  - & ATLAS  \\ 
2016-10-12 & 57673.97 & -1485.2 & - &  $>  18.4 $& ATLAS  \\ 
2016-10-30 & 57691.24 & -1467.9 & $>  19.4 $&  - & ATLAS  \\ 
2016-11-15 & 57707.22 & -1452.0 & - &  $>  18.7 $& ATLAS  \\ 
2016-11-23 & 57715.21 & -1444.0 & $>  19.3 $&  - & ATLAS  \\ 
2017-06-13 & 57917.04 & -1242.1 & - &  $>  19.6 $& ATLAS  \\ 
2017-06-30 & 57934.54 & -1224.6 & - &  $>  19.7 $& ATLAS  \\ 
2017-07-18 & 57952.99 & -1206.2 & - &  $>  19.8 $& ATLAS  \\ 
2017-07-31 & 57965.14 & -1194.0 & - &  $>  19.7 $& ATLAS  \\ 
2017-08-15 & 57980.43 & -1178.8 & - &  $>  19.6 $& ATLAS  \\ 
2017-08-19 & 57984.42 & -1174.8 & $>  20.3 $&  - & ATLAS  \\ 
2017-08-27 & 57992.41 & -1166.8 & - &  $>  19.8 $& ATLAS  \\ 
2017-09-15 & 58011.37 & -1147.8 & - &  $>  19.7 $& ATLAS  \\ 
2017-09-21 & 58017.36 & -1141.8 & $>  20.3 $&  - & ATLAS  \\ 
2017-09-28 & 58024.00 & -1135.2 & - &  $>  19.6 $& ATLAS  \\ 
2017-10-12 & 58038.65 & -1120.5 & - &  $>  19.2 $& ATLAS  \\ 
2017-10-27 & 58053.28 & -1105.9 & - &  $>  18.9 $& ATLAS  \\ 
2017-11-06 & 58063.25 & -1095.9 & - &  $>  19.1 $& ATLAS  \\ 
2017-11-18 & 58075.24 & -1083.9 & - &  $>  19.6 $& ATLAS  \\ 
2017-12-11 & 58098.71 & -1060.5 & - &  $>  19.2 $& ATLAS  \\ 
2018-03-08 & 58185.83 &  -973.3 & $>  20.2 $&  - & ATLAS  \\ 
2018-04-20 & 58228.63 &  -930.6 & - &  $>  18.7 $& ATLAS  \\ 
2018-05-22 & 58260.08 &  -899.1 & - &  $>  19.8 $& ATLAS  \\ 
2018-06-05 & 58274.87 &  -884.3 & - &  $>  19.4 $& ATLAS  \\ 
2018-06-14 & 58283.54 &  -875.6 & $>  20.3 $&  - & ATLAS  \\ 
2018-06-20 & 58289.04 &  -870.1 & - &  $>  19.9 $& ATLAS  \\ 
2018-07-01 & 58300.49 &  -858.7 & - &  $>  19.5 $& ATLAS  \\ 
2018-07-12 & 58311.49 &  -847.7 & $>  20.2 $&  - & ATLAS  \\ 
2018-07-17 & 58316.98 &  -842.2 & - &  $>  19.7 $& ATLAS  \\ 
2018-08-01 & 58331.46 &  -827.7 & - &  $>  19.3 $& ATLAS  \\ 
2018-08-07 & 58337.46 &  -821.7 & $>  20.1 $&  - & ATLAS  \\ 
2018-08-15 & 58345.43 &  -813.8 & $>  18.5 $&  - & ATLAS  \\ 
2018-08-17 & 58347.41 &  -811.8 & - &  $>  19.2 $& ATLAS  \\ 
2018-09-05 & 58366.58 &  -792.6 & - &  $>  19.9 $& ATLAS  \\ 
2018-09-20 & 58381.39 &  -777.8 & - &  $>  19.3 $& ATLAS  \\ 
2018-10-01 & 58392.32 &  -766.9 & - &  $>  19.5 $& ATLAS  \\ 
2018-10-12 & 58403.30 &  -755.9 & $>  20.2 $&  - & ATLAS  \\ 
2018-10-17 & 58408.33 &  -750.8 & - &  $>  19.4 $& ATLAS  \\ 
2018-10-31 & 58422.28 &  -736.9 & - &  $>  19.4 $& ATLAS  \\ 
2018-11-05 & 58427.29 &  -731.9 & $>  20.1 $&  - & ATLAS  \\ 
2018-11-13 & 58435.92 &  -723.3 & - &  $>  19.4 $& ATLAS  \\ 
2018-11-26 & 58448.84 &  -710.3 & - &  $>  19.4 $& ATLAS  \\ 
2018-12-07 & 58459.88 &  -699.3 & - &  $>  19.5 $& ATLAS  \\ 
2018-12-09 & 58461.21 &  -698.0 & $>  19.5 $&  - & ATLAS  \\ 
2018-12-16 & 58468.21 &  -691.0 & - &  $>  19.1 $& ATLAS  \\ 
2019-04-16 & 58589.64 &  -569.5 & - &  $>  18.1 $& ATLAS  \\ 
2019-05-06 & 58609.60 &  -549.6 & $>  20.2 $&  - & ATLAS  \\ 
\hline  
\end{tabular}
\begin{flushleft}
$^a$ Phases are relative to $B$ maximum light, MJD = 59159.18 $\pm$ 0.50.\\ 
\end{flushleft}
\end{table*}

\begin{table*}
\contcaption{(continued)}
\begin{tabular}{@{}ccccccccc@{}}
\hline  
Date & MJD & Phase$^a$ & cyan & orange & Instrument key\\ 
 &  & (days) & (mag)  & (mag) &  \\ 
\hline  
2019-05-14 & 58617.57 &  -541.6 & - &  $>  19.6 $& ATLAS  \\ 
2019-05-24 & 58627.56 &  -531.6 & - &  $>  19.2 $& ATLAS  \\ 
2019-06-01 & 58635.59 &  -523.6 & $>  20.3 $&  - & ATLAS  \\ 
2019-06-07 & 58641.59 &  -517.6 & - &  $>  19.9 $& ATLAS  \\ 
2019-06-17 & 58651.52 &  -507.7 & - &  $>  19.0 $& ATLAS  \\ 
2019-06-29 & 58663.55 &  -495.6 & - &  $>  19.5 $& ATLAS  \\ 
2019-07-05 & 58669.29 &  -489.9 & $>  20.3 $&  - & ATLAS  \\ 
2019-07-10 & 58674.49 &  -484.7 & - &  $>  20.2 $& ATLAS  \\ 
2019-07-26 & 58690.49 &  -468.7 & - &  $>  19.8 $& ATLAS  \\ 
2019-07-29 & 58693.45 &  -465.7 & $>  20.3 $&  - & ATLAS  \\ 
2019-08-06 & 58701.50 &  -457.7 & - &  $>  20.0 $& ATLAS  \\ 
2019-08-10 & 58705.30 &  -453.9 & $>  18.5 $&  - & ATLAS  \\ 
2019-08-22 & 58717.82 &  -441.4 & - &  $>  19.8 $& ATLAS  \\ 
2019-08-30 & 58725.39 &  -433.8 & $>  20.3 $&  - & ATLAS  \\ 
2019-09-20 & 58746.02 &  -413.2 & - &  $>  19.6 $& ATLAS  \\ 
2019-09-25 & 58751.35 &  -407.8 & $>  20.2 $&  - & ATLAS  \\ 
2019-10-02 & 58758.65 &  -400.5 & - &  $>  19.8 $& ATLAS  \\ 
2019-10-17 & 58773.69 &  -385.5 & - &  $>  19.7 $& ATLAS  \\ 
2019-10-27 & 58783.28 &  -375.9 & $>  20.2 $&  - & ATLAS  \\ 
2019-10-29 & 58785.31 &  -373.9 & - &  $>  19.9 $& ATLAS  \\ 
2019-11-11 & 58798.27 &  -360.9 & - &  $>  19.3 $& ATLAS  \\ 
2019-11-30 & 58817.21 &  -342.0 & $>  19.4 $&  - & ATLAS  \\ 
2019-12-01 & 58818.20 &  -341.0 & - &  $>  19.3 $& ATLAS  \\ 
2019-12-10 & 58827.20 &  -332.0 & - &  $>  19.3 $& ATLAS  \\ 
2019-12-17 & 58834.20 &  -325.0 & $>  19.1 $&  - & ATLAS  \\ 
2020-04-16 & 58955.61 &  -203.6 & - &  $>  19.5 $& ATLAS  \\ 
2020-04-26 & 58965.60 &  -193.6 & $>  20.2 $&  - & ATLAS  \\ 
2020-04-30 & 58969.11 &  -190.1 & - &  $>  19.5 $& ATLAS  \\ 
2020-05-15 & 58984.08 &  -175.1 & - &  $>  19.6 $& ATLAS  \\ 
2020-05-26 & 58995.57 &  -163.6 & - &  $>  19.8 $& ATLAS  \\ 
2020-05-27 & 58996.26 &  -162.9 & $>  20.2 $&  - & ATLAS  \\ 
2020-06-06 & 59006.87 &  -152.3 & - &  $>  19.4 $& ATLAS  \\ 
2020-06-18 & 59018.52 &  -140.7 & - &  $>  20.0 $& ATLAS  \\ 
2020-06-22 & 59022.83 &  -136.3 & $>  20.1 $&  - & ATLAS  \\ 
2020-06-30 & 59030.71 &  -128.5 & - &  $>  19.8 $& ATLAS  \\ 
2020-07-13 & 59043.02 &  -116.2 & - &  $>  19.9 $& ATLAS  \\ 
2020-07-21 & 59051.55 &  -107.6 & $>  20.3 $&  - & ATLAS  \\ 
2020-07-26 & 59056.95 &  -102.2 & - &  $>  19.9 $& ATLAS  \\ 
2020-08-10 & 59071.46 &   -87.7 & - &  $>  19.9 $& ATLAS  \\ 
2020-08-16 & 59077.40 &   -81.8 & $>  20.1 $&  - & ATLAS  \\ 
2020-08-20 & 59081.41 &   -77.8 & - &  $>  20.1 $& ATLAS  \\ 
2020-09-05 & 59097.39 &   -61.8 & - &  $>  19.8 $& ATLAS  \\ 
2020-09-07 & 59099.37 &   -59.8 & - &  19.28 (0.43) & ATLAS  \\ 
2020-09-09 & 59101.35 &   -57.8 & 19.64 (0.36) &  19.26 (0.18) & ATLAS  \\ 
2020-09-10 & 59102.37 &   -56.8 & 19.72 (0.37) &  - & ATLAS  \\ 
2020-09-11 & 59103.35 &   -55.8 & - &  19.28 (0.20) & ATLAS  \\ 
2020-09-13 & 59105.36 &   -53.8 & 19.72 (0.26) &  - & ATLAS  \\ 
2020-09-15 & 59107.34 &   -51.8 & - &  19.09 (0.18) & ATLAS  \\ 
2020-09-17 & 59109.35 &   -49.8 & 19.06 (0.18) &  - & ATLAS  \\ 
2020-09-19 & 59111.33 &   -47.8 & - &  18.64 (0.14) & ATLAS  \\ 
2020-09-21 & 59113.33 &   -45.8 & 18.81 (0.27) &  - & ATLAS  \\ 
2020-09-23 & 59115.30 &   -43.9 & - &  18.39 (0.11) & ATLAS  \\ 
2020-09-25 & 59117.28 &   -41.9 & - &  18.39 (0.17) & ATLAS  \\ 
2020-10-01 & 59123.35 &   -35.8 & - &  18.05 (0.13) & ATLAS  \\ 
2020-10-05 & 59127.28 &   -31.9 & - &  17.85 (0.08) & ATLAS  \\ 
2020-10-07 & 59129.31 &   -29.9 & - &  17.67 (0.06) & ATLAS  \\ 
2020-10-09 & 59131.33 &   -27.9 & - &  17.56 (0.05) & ATLAS  \\ 
2020-10-11 & 59133.32 &   -25.9 & 17.66 (0.05) &  - & ATLAS  \\ 
\hline  
\end{tabular}
\begin{flushleft}
$^a$ Phases are relative to $B$ maximum light, MJD = 59159.18 $\pm$ 0.50.\\ 
\end{flushleft}
\end{table*}

\begin{table*}
\contcaption{(continued)}
\begin{tabular}{@{}ccccccccc@{}}
\hline  
Date & MJD & Phase$^a$ & cyan & orange & Instrument key\\ 
 &  & (days) & (mag)  & (mag) &  \\ 
\hline  
2020-10-13 & 59135.28 &   -23.9 & - &  17.54 (0.06) & ATLAS  \\ 
2020-10-15 & 59137.26 &   -21.9 & 17.42 (0.04) &  - & ATLAS  \\ 
2020-10-17 & 59139.30 &   -19.9 & - &  17.29 (0.04) & ATLAS  \\ 
2020-10-19 & 59141.33 &   -17.9 & 17.55 (0.34) &  - & ATLAS  \\ 
2020-10-21 & 59143.26 &   -15.9 & - &  17.12 (0.05) & ATLAS  \\ 
2020-10-23 & 59145.25 &   -13.9 & - &  17.13 (0.05) & ATLAS  \\ 
2020-10-29 & 59151.27 &    -7.9 & - &  17.06 (0.05) & ATLAS  \\ 
2020-11-04 & 59157.24 &    -1.9 & - &  17.00 (0.07) & ATLAS  \\ 
2020-11-08 & 59161.24 &     2.1 & 17.07 (0.04) &  - & ATLAS  \\ 
2020-11-10 & 59163.24 &     4.1 & - &  17.01 (0.03) & ATLAS  \\ 
2020-11-16 & 59169.23 &    10.0 & 17.03 (0.03) &  - & ATLAS  \\ 
2020-11-28 & 59181.22 &    22.0 & - &  17.09 (0.07) & ATLAS  \\ 
2020-12-02 & 59185.20 &    26.0 & - &  17.11 (0.05) & ATLAS  \\ 
2020-12-04 & 59187.21 &    28.0 & - &  17.12 (0.04) & ATLAS  \\ 
2020-12-06 & 59189.19 &    30.0 & 17.24 (0.06) &  - & ATLAS  \\ 
2020-12-08 & 59191.19 &    32.0 & - &  17.14 (0.07) & ATLAS  \\ 
2020-12-12 & 59195.19 &    36.0 & 17.29 (0.06) &  - & ATLAS  \\ 
2021-04-15 & 59319.62 &   160.4 & $>  20.2 $&  - & ATLAS  \\ 
2021-04-22 & 59326.60 &   167.4 & - &  $>  19.5 $& ATLAS  \\ 
2021-04-30 & 59334.61 &   175.4 & - &  $>  19.1 $& ATLAS  \\ 
2021-05-15 & 59349.56 &   190.4 & $>  20.1 $&  $>  20.0 $& ATLAS  \\ 
2021-05-15 & 59349.60 &   190.4 & $>  20.1 $&  $>  20.0 $& ATLAS  \\ 
\hline  
\end{tabular}
\begin{flushleft}
$^a$ Phases are relative to $B$ maximum light, MJD = 59159.18 $\pm$ 0.50.\\ 
\end{flushleft}
\end{table*}

\begin{table*}
  \caption{Log of spectroscopic observations of \pvb.}
  \label{table_spec}
  \begin{tabular}{@{}ccccccc@{}}
  \hline
  Date & MJD & Phase$^a$ & Instrumental set-up & Grism/grating + slit & Spectral range & Resolution \\
 & ($-$2,400,000.00) & & & & (\AA) & (\AA) \\
 \hline
20201012 & 59134.84  &  -24.3 & TNG+LRS       &    LR-B+1$\farcs$5  &  3450-8000 & 14.5 \\			    
20201015 & 59137.88  &  -21.3  &  NOT+ALFOSC    &    gr4+1$\farcs$0 & 3500-9680 & 14 \\			    
20201026 & 59148.86  &  -10.3  &  NOT+ALFOSC    &    gr4+1$\farcs$0 & 3500-9680 & 14 \\			    
20201027 & 59149.08  &  -10.1  &  NTT+EFOSC2    &    gr11+gr16+1$\farcs$0  &  3380-7470  &  14 \\		    
20201030 & 59152.08  &  -7.1	  &  NTT+EFOSC2  &    gr11+gr16+1$\farcs$0  &  3360-10000  &  14 \\		    
20201106 & 59159.02  &  -0.2	  &  NTT+EFOSC2  &    gr11+1$\farcs$0  &  3360-10000  &  14 \\		    
20201107 & 59160.85  &  1.7	  &  GTC+OSIRIS  &    R1000B+1$\farcs$0  &  5580-7685  &  7 \\	    
20201107 & 59160.86  &  1.7	  &  GTC+OSIRIS  &    R2500R+R2500I+1$\farcs$0  &  7330-10155  &  3.5 \\	    
20201117 & 59170.06  &  10.9	  &  NTT+EFOSC2  &    gr11+gr16+1$\farcs$0  &  3355-10000  &  14 \\		    
20201123 & 59176.03  &  16.9	  &  NTT+EFOSC2  &    gr11+gr16+1$\farcs$0  &  3355-10000  &  14 \\		    
20201124 & 59177.81  &  18.6	  &  GTC+OSIRIS  &    R2500R+1$\farcs$0  &  5580-7685  &  3.5  \\	    
20201207 & 59190.05  &  30.9	  &  NTT+EFOSC2  &    gr11+1$\farcs$0  &  3400-7470  &  14 \\			    
20201208 & 59191.03  &  31.9	  &  NTT+EFOSC2  &    gr16+1$\farcs$0  &  6000-10000  & 14 \\                     
\hline
\end{tabular}
\begin{flushleft}
$^a$ Phases are relative to $B$ maximum light, MJD = 59159.18 $\pm$ 0.50.\\ \end{flushleft}
\end{table*}

\begin{table*}
\caption{$UVW2,UVM2,UVW1$ ({\sc Vega mag}) photometry of \pvb.}
\label{table_UVph}
\begin{tabular}{@{}ccccccc@{}}
\hline  
Date & MJD & Phase$^a$ & UVW2 & UVM2 & UVW1 & Instrument key\\ 
 &  & (days) & (mag) & (mag)  & (mag) &  \\ 
\hline  
2019-10-04 & 58760.44 &  -398.7 & $>  19.5 $&  $>  19.6 $&  $>  20.2 $& UVOT \\ 
2020-10-20 & 59142.76 &   -16.4 & 17.99 (0.14) &  17.64 (0.09) &  17.27 (0.12) & UVOT  \\ 
2020-10-20 & 59142.96 &   -16.2 & 17.54 (0.12) &  17.81 (0.10) &  17.25 (0.12) & UVOT  \\ 
2020-11-18 & 59171.35 &    12.2 & 17.50 (0.10) &  17.29 (0.08) &  16.97 (0.08) & UVOT  \\ 
\hline  
\end{tabular}
\begin{flushleft}
$^a$ Phases are relative to $B$ maximum light, MJD = 59159.18 $\pm$ 0.50.\\ 
\end{flushleft}
\end{table*}

\end{appendix}
\end{document}